\documentclass[journal]{IEEEtran}

\usepackage{xspace,amsmath,amssymb,amsfonts,epsfig,subfigure,syntonly}
\usepackage{cite,bm,color,url,textcomp,empheq,boxedminipage}
\usepackage{algorithmicx,algorithm}
\usepackage{epstopdf,makecell}
\usepackage{empheq}
\usepackage{pifont}
\usepackage{soul}

\usepackage{graphicx,graphics}  
\usepackage{multirow,multicol}
\usepackage{psfrag}    
\usepackage{stfloats}
\usepackage{url}

\newtheorem{lemma}{\underline{Lemma}}[section]

\newtheorem{proposition}{\underline{Proposition}}[section]

\newtheorem{remark}{\underline{Remark}}[section]

\long\def\symbolfootnote[#1]#2{\begingroup
\def\thefootnote{\fnsymbol{footnote}}
\footnote[#1]{#2}\endgroup}

\psfull

\allowdisplaybreaks[4]

\begin{document}
\title{Throughput Maximization for UAV-Enabled Wireless Powered Communication Networks}


\author{Lifeng~Xie,~\IEEEmembership{Student Member,~IEEE,} Jie~Xu,~\IEEEmembership{Member,~IEEE,} and Rui~Zhang,~\IEEEmembership{Fellow,~IEEE}\\
\thanks{Part of this paper has been presented at the IEEE Vehicular Technology Conference (VTC2018-spring), Porto, Portugal, June 3-6, 2018\cite{conference}. {\it (Corresponding author: Jie Xu.)}}
\thanks{L. Xie and J. Xu are with the School of Information Engineering, Guangdong University of Technology, Guangzhou, China (e-mail: lifengxie22039@gmail.com, jiexu@gdut.edu.cn).}
\thanks{R. Zhang is with the Department of Electrical and Computer Engineering, National University of Singapore, Singapore (e-mail: elezhang@nus.edu.sg).}
\thanks{Copyright (c) 2012 IEEE. Personal use of this material is permitted. However, permission to use this material for any other purposes must be obtained from the IEEE by sending a request to pubs-permissions@ieee.org.}
}

%

\setlength\abovedisplayskip{1pt}
\setlength\belowdisplayskip{1pt}
\maketitle
\begin{abstract}
This paper studies an unmanned aerial vehicle (UAV)-enabled wireless powered communication network (WPCN), in which a UAV is dispatched as a mobile access point (AP) to serve a set of ground users periodically. The UAV employs the radio frequency (RF) wireless power transfer (WPT) to charge the users in the downlink, and the users use the harvested RF energy to send independent information to the UAV in the uplink. Unlike the conventional WPCN with fixed APs, the UAV-enabled WPCN can exploit the mobility of the UAV via trajectory design, jointly with the wireless resource allocation optimization, to maximize the system throughput. In particular, we aim to maximize the uplink common (minimum) throughput among all ground users over a finite UAV's flight period, subject to its maximum speed constraint and the users' energy neutrality constraints. The resulted problem is non-convex and thus difficult to be solved optimally. To tackle this challenge, we first consider an ideal case without the UAV's maximum speed constraint, and obtain the optimal solution to the relaxed problem. The optimal solution shows that the UAV should successively hover above a finite number of ground locations for downlink WPT, as well as above each of the ground users for uplink communication. Next, we consider the general problem with the UAV's maximum speed constraint. Based on the above multi-location-hovering solution, we first propose an efficient successive hover-and-fly trajectory design, jointly with the downlink and uplink wireless resource allocation, and then propose a locally optimal solution by applying the techniques of alternating optimization and successive convex programming (SCP). Numerical results show that the proposed UAV-enabled WPCN achieves significant throughput gains over the conventional WPCN with fixed-location AP.
\end{abstract}



\section{Introduction}
\IEEEPARstart{R}{adio} frequency (RF) wireless power transfer (WPT) has emerged as a promising solution to provide convenient and reliable energy supply to low-power Internet-of-things (IoT) devices such as sensors and RF identification (RFID) tags\cite{WPTsurvey,YongSurvey,More22,More33}. Compared to the near-field WPT based on inductive coupling or magnetic resonant coupling, the far-field WPT via RF radiation is able to operate over a much longer range and charge multiple wireless devices (WDs) simultaneously even when they are moving and densely deployed, and yet with transceivers of significantly reduced form factor. There are in general two major applications of RF WPT in wireless communications, namely simultaneous wireless information and power transfer (SWIPT) and  wireless powered communication network (WPCN), which unify the WPT and wireless information transfer (WIT) in a joint design framework over the same (downlink for both WPT and WIT) and opposite (downlink for WPT and uplink for WIT) transmission directions, respectively\cite{Bi2015,BiBi}. In particular, WPCN enables dedicated wireless charging and information collection for massive IoT devices, thus significantly enhances the operation range and throughput of traditional backscattering-based wireless communications. However, in conventional WPCNs, access points (APs) are usually deployed at fixed locations, which cannot be changed once deployed \cite{Rui2013WPCN,Placement,WangFeng}.

The conventional WPCN with fixed APs faces several challenges. First, due to the severe propagation loss of RF signals over distance, the end-to-end WPT efficiency is generally low when the distance from the AP to a WD becomes large. Next, the conventional WPCN suffers from the so-called ``doubly near-far'' problem\cite{Rui2013WPCN}, i.e., far-apart WDs from the AP receive lower RF energy in the downlink WPT, but they need to use higher transmit power in the uplink WIT to achieve the same rate as nearby WDs. The doubly near-far problem result in a severe user fairness issue among WDs when they are geographically distributed over a large area. To overcome the above issues, various approaches have been proposed in the literature, such as adaptive time and power allocation \cite{Rui2013WPCN}, multi-antenna beamforming\cite{Xu2014A,LLiuantenna,Xu2014B,JXUMIMO,DUAN}, and user cooperation \cite{Usercooperation,WPCC}. However, all these prior works focused on the wireless resource allocation designs to enhance the WPT/WIT performances of WPCNs with fixed APs. By contrast, in this paper we propose an alternative solution based on a new unmanned aerial vehicle (UAV)-enabled WPCN architecture with UAVs employed as mobile APs.
\begin{figure}
\setlength{\abovecaptionskip}{-0.mm}
\setlength{\belowcaptionskip}{-0.mm}
\centering
    \includegraphics[width=7cm]{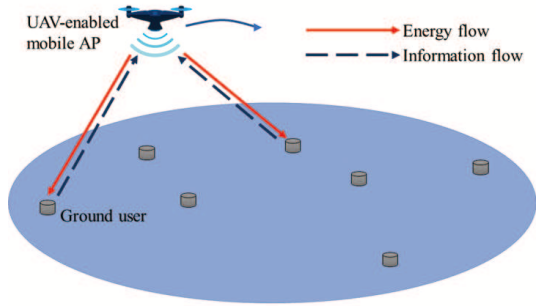}
\caption{Illustration of a UAV-enabled WPCN.} \label{fig:1}
\end{figure}

UAVs have found abundant applications such as for cargo delivery, aerial surveillance, filming, and industrial IoT. Recently, UAVs-enabled/aided  wireless communications have attracted substantial research interests, due to their advantages in flexible deployment, strong line-of-sight (LoS) channels with ground users, and controllable mobility\cite{ZYwireless}. For example, UAVs can be utilized as mobile relays to help information exchange between far-apart ground users \cite{zeng2016throughput}, or as mobile base stations (BSs) to help enhance the wireless coverage and/or the network capacity for ground mobile users \cite{JLyu,C.Zhan,RIB,MMoza,WuQing,Q.WU,Y.WU,DavidGesbert,walidsaad}. Furthermore, UAV-enabled WPT has been proposed in \cite{JieXuWPT,Jiewpt}, in which UAVs are used as mobile energy transmitters to charge low-power WDs on the ground. By exploiting its fully controllable mobility, the UAV can properly adjust its location over time (a.k.a. trajectory) to reduce the distances with target ground users, thus improving the efficiency for both WIT and WPT.

Motivated by the UAV-enabled wireless communications as well as WPT, this paper pursues a unified study on both of them in a UAV-enabled WPCN as shown in Fig. 1. Specifically, a UAV following a pre-designed periodic trajectory is dispatched as a mobile AP to charge a set of ground users in the downlink via WPT, and the users use the harvested RF energy to send independent information to the UAV in the uplink. We investigate how to optimally exploit the UAV mobility via trajectory design, jointly with the wireless resource allocation, to maximize the uplink data throughput of the multiuser WPCN in a fair manner. To this end, we maximize the uplink common (minimum) throughput among all ground users over a given UAV's flight period, by optimizing its trajectory, jointly with the downlink and uplink transmission resource allocations for WPT and WIT, respectively, subject to the UAV's maximum speed and the users' energy neutrality constraints. However, due to the complex data throughput and harvested energy functions in terms of coupled UAV trajectory and resource allocation design variables, our formulated problem is non-convex and thus difficult to be solved optimally.

To tackle this difficulty, we first consider an ideal case without considering the UAV's maximum speed constraint. We show that the strong duality holds between this problem and its Lagrange dual problem, and thus it can be solved optimally via the Lagrange dual method. The optimal solution shows that the UAV should successively hover above a finite number of ground locations for downlink WPT, as well as above each of the ground users for uplink WIT, with the optimal hovering duration and wireless resource allocation for each location. Next, we address the general problem with the UAV's maximum speed constraint considered. Based on the multi-location-hovering solution to the above relaxed problem, we first propose a heuristic {\it{successive hover-and-fly trajectory}} design, jointly with the downlink and uplink resource allocations, to find an efficient suboptimal solution. The proposed solution is also shown to be asymptotically optimal as the UAV's flight period becomes infinitely large. In addition, we further propose an alternating optimization based algorithm to obtain a locally optimal solution, which optimizes the wireless resource allocations and the UAV trajectory in an alternating manner, via convex optimization and successive convex programming (SCP) techniques, respectively. By employing the successive hover-and-fly trajectory as the UAV initial trajectory, the alternating-optimization-based algorithm iteratively refines the wireless resource allocations and the UAV trajectory to improve the uplink common throughput of all ground users until convergence. Finally, we present numerical results to validate the performance of our proposed UAV-enabled WPCN. It is shown that the joint trajectory and wireless resource allocation design significantly improves the uplink common throughput, as compared to the conventional WPCN with the AP at a fixed location.

It is worth noting that there is another line of related research that employs ground moving vehicles to wirelessly charge and/or collect information from ground sensors, e.g., \cite{L.Xie,JMchen,Y.Shu,More11}. However, different from the UAV that can freely fly in the three-dimensional (3D) airspace, the ground vehicle can only move following a constrained path in the two-dimensional (2D) plane. Furthermore, unlike UAVs that have strong LoS links with ground users, the wireless channels from ground vehicles to users usually suffer from severe fading, thus limiting the performance for both WIT and WPT. As such, the joint trajectory design and wireless resource allocation in the UAV-enabled WPCN is a new study different from that with ground moving vehicles, which has not been investigated in the literature to our best knowledge.

The remainder of this paper is organized as follows. Section \ref{sec:system} presents the system model of the UAV-enabled WPCN, and formulates the uplink common throughput maximization problem of our interest. Section \ref{sec:optimal} considers an ideal case without the UAV's maximum speed constraint and presents the optimal solution to the relaxed problem. Section \ref{sec:P1} and Section V present two efficient solutions to the general problem with the UAV's maximum speed constraint considered. Section \ref{sec:simulation} provides numerical results to validate the effectiveness of our proposed designs. Finally, Section \ref{sec:conclusions} concludes the paper.

\section{System Model and Problem Formulation}\label{sec:system}\vspace{-0cm}

As shown in Fig. \ref{fig:1}, we consider a UAV-enabled WPCN, in which a UAV is dispatched to periodically charge a set $\mathcal K \triangleq \{1,\ldots,K\}$ of ground users via WPT in the downlink, and each user $k\in \mathcal K$ uses its harvested energy to send independent information to the UAV in the uplink. Suppose that each user $k\in\mathcal{K}$ is at a fixed location $(x_k,y_k,0)$ on the ground in a 3D Cartesian coordinate system, where $\bold{w}_k =(x_k,y_k)$ is defined as the horizontal coordinate of user $k$. The users' locations are assumed to be {\it a-priori} known by the UAV for its trajectory design and transmission resource allocation.

We focus on one particular flight period of the UAV, denoted by $\mathcal T \triangleq (0,T]$ with finite duration $T$ in second (s), in which the UAV flies horizontally at a fixed altitude $H>0$ in meter (m). At any given time instant $t\in \mathcal T$, let ${\bold{q}}(t)=(x(t),y(t))$ denote the location of the UAV projected on the horizontal plane. Accordingly, the distance between the UAV and each user $k \in \mathcal K$ is given by
\begin{align}
d_k(\bold{q}(t))&=\sqrt{\lVert\bold{q}(t)-\bold{w}_k\rVert^2+H^2},
\end{align}
where $\| \cdot \|$ denotes the Euclidean norm of a vector. By denoting the UAV's maximum speed as $V_{\max}$ in m/s, we have $
\sqrt{\dot{x}^2(t)+\dot{y}^2(t)}\le V_{\max}, \forall t\in\mathcal T$,
where $\dot x(t)$ and $\dot y(t)$ denote the first-derivatives of $x(t)$ and $y(t)$ with respect to $t$, respectively. Note that we assume the UAV can freely choose its initial location $\bold{q}(0)$ and final location $\bold{q}(T)$ for performance optimization.

We consider that the wireless channels between the UAV and the $K$ ground users are dominated by LoS links. 
In this case, the free-space path loss model can be practically assumed, 
similarly as in\cite{zeng2016throughput,JieXuWPT}. Accordingly, the channel power gain between the UAV and user $k\in\mathcal K$ at time instant $t\in\mathcal T$ is given by
\begin{align}
h_k(\bold{q}(t))&=\beta_{\rm 0}d^{-2}_k(\bold{q}(t))=\frac{\beta_{\rm 0}}{\lVert\bold{q}(t)-\bold{w}_k\rVert^2+H^2},
\end{align}
where $\beta_{\rm 0}$ denotes the channel power gain at a reference distance of $d_{\rm 0}=1$ m.

We consider a time-division multiple access (TDMA) transmission protocol, in which the downlink WPT for all users and the uplink WIT of different users are implemented in the same frequency band but over orthogonal time instants. At any time instant $t\in\mathcal T$, we use the indicators $\rho_k(t) \in \{0,1\}, \forall k\in\{0\} \cup \mathcal K$, to denote the transmission mode. We use $\rho_0(t) = 1$ and $\rho_k(t) = 0, \forall k\in\mathcal K$, to indicate the downlink WPT mode, in which the UAV transmits RF energy to charge the $K$ users simultaneously; while we use $\rho_k(t) = 1, k\in\mathcal K$, and $\rho_j(t) = 0, \forall j\in \{0\} \cup\mathcal K, j\neq k$, to represent the uplink WIT mode for user $k$, in which user $k$ sends its information to the UAV by using its harvested energy. As the TDMA protocol is employed, it follows that $\sum_{k=0}^{K} \rho_k(t) = 1, \forall t\in\mathcal T$.

First, consider the downlink WPT mode at time instant $t\in\mathcal T$, in which $\rho_0(t) = 1$, and $\rho_k(t) = 0, \forall k\in\mathcal K$. Suppose that the UAV adopts a constant transmit power $P$ in the downlink WPT mode. Accordingly, the harvested power at each user $k\in\mathcal K$ is given by
\begin{align}
E_k(\rho_0(t),\bold{q}(t))&=\eta P\rho_0(t)h_k(\bold{q}(t))\nonumber\\
&=\frac{\eta P\beta_{\rm 0}\rho_0(t) }{\lVert\bold{q}(t)-\bold{w}_k\rVert^2+H^2},\label{eqn:linear:model}
\end{align}
where $0<\eta\le 1$ denotes the RF-to-direct current (DC) energy conversion efficiency at the energy harvester of each user.\footnote{Note that in practice, the RF-to-DC energy conversion efficiency is generally non-linear and depends on the received RF power level \cite{Boshko} and signal waveform \cite{Clerck}. For the purpose of exposition, in (\ref{eqn:linear:model}) we consider a simplified constant RF-to-DC energy conversion efficiency by assuming that each receiver operates at the linear regime for RF-to-DC conversion. Nevertheless, the design principles in this paper are also extendable to the scenario with non-linear RF-to-DC energy conversion efficiency, which, however, is left for future work. } 
Therefore, the total harvested energy at user $k$ over each period of duration $T$ is given by
\begin{align}\label{havestedenergy}
\hat E_k(\{\rho_0(t),\bold{q}(t)\}) = \int_{0}^T E_k(\rho_0(t),\bold{q}(t))\text{d}t.
\end{align}

Next, consider the WIT mode for user $k\in\mathcal K$ at time instant $t \in\mathcal T$ with $\rho_k(t) = 1$, and $\rho_j(t) = 0, \forall j\in\{0\}\cup\mathcal K, j\neq k$. Let $Q_k(t)$ denote the transmit power of user $k$ for the uplink WIT to the UAV. Accordingly, the achievable data rate from user $k$ to the UAV in bits/second/Hertz (bps/Hz) at time instant $t\in\mathcal T$ is given by
\begin{align}
r_k(\rho_k(t),&\bold{q}(t),Q_k(t))=\rho_k(t){\rm{log_2}}\left(1+\frac{Q_k(t)h_k(\bold{q}(t))}{\sigma^2}\right)\nonumber\\
&~~=\rho_k(t){\rm{log_2}}\left(1+\frac{Q_k(t)\gamma}{\lVert\bold{q}(t)-\bold{w}_k\rVert^2+H^2}\right),
\end{align}
where $\sigma^2$ denotes the noise power at the information receiver of the UAV, and $\gamma\triangleq\beta_{\rm 0}/\sigma^2$ is the reference signal-to-noise ratio (SNR).

Therefore, the average achievable rate or throughput of user $k$ over each period in bps/Hz is given by
\begin{align}
R_k(\{\rho_k(t),\bold{q}(t),Q_k(t)\})=\frac{1}{T}\int_0^{T}r_k(\rho_k(t),\bold{q}(t),Q_k(t))\text{d}t.
\end{align}
Note that for the purpose of exposition, we consider that the energy consumption of each ground user is mainly due to the transmit power for its uplink WIT. In this case, the total energy consumption at user $k\in\mathcal K$ is
\begin{align}\label{energyconsumption}
\hat{Q}_k(\{\rho_k(t),Q_k(t)\}) = \int_0^T \rho_k(t)Q_k(t) \text{d}t.
\end{align}
In order to achieve the self-sustainable operation for the WPCN, we consider the energy neutrality constraint at each user $k$, such that the user's energy consumption for uplink WIT (i.e., $\hat{Q}_k(\{\rho_k(t),Q_k(t)\})$ in (\ref{energyconsumption})) cannot exceed the energy harvested from the downlink WPT (i.e., $\hat E_k(\{\rho_0(t),\bold{q}(t)\})$ in (\ref{havestedenergy})) in each period.\footnote{ Note that we assume that at the beginning of each period, each user has sufficient energy in its storage and the storage has sufficiently large capacity. In this case, as long as the energy neutrality constraints are satisfied over each period, the more stringent energy causality constraints in energy harvesting wireless communications \cite{H.Li} will be automatically ensured, and thus each of the $K$ users can sustain its operation without energy outage.}
As a result, we have the following energy neutrality constraints for the $K$ users,
\begin{align}\label{energyconstraint}
\int_0^T \rho_k(t) Q_k(t) \text{d}t \le \int_0^T E_k(\rho_0(t),\bold{q}(t))\text{d}t, \forall k\in\mathcal K.
\end{align}

In this work, our objective is to maximize the uplink common throughput among all users (i.e., $\min_{k\in\mathcal{K}}R_k(\{\rho_k(t),\bold{q}(t),Q_k(t)\})$) subject to the UAV's maximum speed constraint and the $K$ users' energy neutrality constraints. The decision variables include the UAV trajectory $\{\bold{q}(t)\}$, the transmission mode $\{\rho_k(t)\}$, and the transmit power $\{Q_k(t)\}$ for uplink WIT. As a result, the problem is formulated as
\begin{align}
\text{(P1)}&:\max_{\{\rho_k(t),Q_k(t),\bold{q}(t)\}} ~ \min_{k\in\mathcal{K}}R_k(\{\rho_k(t),\bold{q}(t),Q_k(t)\})\nonumber\\
\mathrm{s.t.}&\int_0^T \rho_k(t) Q_k(t) \text{d}t \le \int_0^T E_k(\rho_0(t),\bold{q}(t))\text{d}t, \forall k\in\mathcal K\label{eneryconstraints}\\
~&Q_k(t) \ge 0,\forall k \in\mathcal{K}, t\in\mathcal T\label{Qk}\\
~&\rho_k(t)\in\{0,1\},\forall k \in\{0\}\cup\mathcal{K}, t\in\mathcal T\label{rho}\\
~&\sum_{k=0}^K \rho_k(t)= 1,\forall t\in\mathcal T\label{rhosum}\\
~&\sqrt{\dot{x}^2(t)+\dot{y}^2(t)}\le V_{\max}, \forall t\in\mathcal T,\label{speed}
\end{align}
where (\ref{speed}) denotes the UAV's maximum speed constraint.

It is observed that for problem (P1), the objective function is non-concave and constraints (\ref{eneryconstraints}) and (\ref{rho}) are non-convex, due to the complicated rate and energy functions with respect to coupled variables $\rho_k(t)$, $\bold{q}(t)$, and $Q_k(t)$, as well as the binary constraints on $\rho_k(t)$'s. Therefore, (P1) is a non-convex optimization problem. Furthermore, (P1) contains an infinite number of optimization variables over continuous time. For these reasons, problem (P1) is difficult to be solved optimally.  To tackle this problem, in Section \ref{sec:optimal} we first consider an ideal case by ignoring the  UAV's maximum speed constraint in (\ref{speed}), and solve the relaxed problem of (P1) as follows:
\begin{align}
\text{(P2)}:\max_{\substack{\{\rho_k(t),Q_k(t),\bold{q}(t)\}}} &~ \min_{k\in\mathcal{K}}R_k(\{\rho_k(t),\bold{q}(t),Q_k(t)\})\nonumber\\
\mathrm{s.t.}~~~~~~&\text{(\ref{eneryconstraints}), (\ref{Qk}), (\ref{rho}) and (\ref{rhosum}).}\nonumber
\end{align}
Note that problem (P2) also corresponds to the practical scenario when the UAV's flight duration $T$ is sufficiently large for any given finite $V_{\text{max}}$ such that the flying time of the UAV becomes negligible as compared to its hovering time (see Section III for details). In Section IV and Section V, we propose efficient algorithms to solve the general problem (P1) with the UAV's maximum speed constraint based on the optimal solution obtained for the relaxed problem (P2).

\section{Optimal Solution to Problem (P2)}\label{sec:optimal}\vspace{-0em}
In this section, we consider problem (P2). By introducing an auxiliary variable $R$, problem (P2) can be equivalently expressed as
\begin{align}
\text{(P2.1)}:&\max_{\substack{\{\rho_k(t),Q_k(t),\bold{q}(t)\}},R} ~ R\nonumber\\
\mathrm{s.t.}&~\frac{1}{T}\int_{0}^T r_k(\rho_k(t),\bold{q}(t),Q_k(t))\text{d}t\ge R,\forall k \in\mathcal{K}\label{R}\\
~&\text{(\ref{eneryconstraints}),  (\ref{Qk}), (\ref{rho}), and (\ref{rhosum}).}\nonumber
\end{align}
Although problem (P2.1) is still non-convex, one can easily show that it satisfies the so-called time-sharing condition in \cite{dual}. Therefore, the strong duality holds between (P2.1) and its Lagrange dual problem. As a result, we can optimally solve (P2.1) by using the Lagrange dual method.

Let $\lambda_k\ge0$ and $\mu_k\ge0$, $k\in\mathcal{K}$, denote the dual variables associated with the $k$-th constraints in (\ref{R}) and (\ref{eneryconstraints}), respectively. For notational convenience, we define $\boldsymbol{\lambda}\triangleq[\lambda_1,\cdots,\lambda_K]$ and $\boldsymbol{\mu}\triangleq[\mu_1,\cdots,\mu_K]$. The partial Lagrangian of (P2.1) is
\begin{align}
&\mathcal{L}(\{\rho_k(t),Q_k(t),\bold{q}(t)\},R,\boldsymbol{\lambda},\boldsymbol{\mu})\nonumber\\
&=\left(1-\sum_{k=1}^K\lambda_k\right) R+\sum_{k=1}^K\frac{\lambda_k}{T}\int_{0}^T r_k(\rho_k(t),\bold{q}(t),Q_k(t))\text{d}t\nonumber\\
&+\sum_{k=1}^K \mu_k\bigg(\int_{0}^T E(\rho_0(t),\bold{q}(t))\text{d}t-\int_0^{T}\rho_k(t)Q_k(t)\text{d}t\bigg).
\end{align}
The Lagrange dual function of (P2.1) is
\begin{align}
&g(\boldsymbol{\lambda},\boldsymbol{\mu})=\max_{\substack{\{\bold{q}(t),Q_k(t),\rho_k(t)\},R}}\mathcal{L}(\{\rho_k(t),Q_k(t),\bold{q}(t)\},R,\boldsymbol{\lambda},\boldsymbol{\mu})\nonumber\\
&~~~~~~~~~~~~~~~~~~~~~~\mathrm{s.t.}~~~~~~{\text{ (\ref{Qk}), (\ref{rho}), and (\ref{rhosum}).}}\label{g}
\end{align}
\begin{lemma}
In order for the dual function $g(\boldsymbol{\lambda},\boldsymbol{\mu})$ to be upper bounded from above (i.e., $g(\boldsymbol{\lambda},\boldsymbol{\mu})<\infty$), it must hold that $\sum_{k=1}^K \lambda_k=1$.
\end{lemma}
\begin{IEEEproof}
Suppose that $\sum_{k=1}^K \lambda_k>1$ (or $\sum_{k=1}^K \lambda_k<1$). Then by setting $R\rightarrow-\infty$ (or $R\rightarrow\infty$), we have $g(\boldsymbol{\lambda},\boldsymbol{\mu})\rightarrow\infty$. Therefore, $\sum_{k=1}^K \lambda_k = 1$ must hold in order for $g(\boldsymbol{\lambda},\boldsymbol{\mu})$ to be bounded from above, and this lemma is proved.
\end{IEEEproof}

Based on Lemma 3.1, the dual problem of problem (P2.1) is given by
\begin{align}
{\text{(D2.1)}}:\min_{\boldsymbol{\lambda},\boldsymbol{\mu}}~& g(\boldsymbol{\lambda},\boldsymbol{\mu})\nonumber\\
\mathrm{s.t.}~&\sum_{k=1}^K \lambda_k =1\nonumber\\
~&\lambda_k\ge 0,\mu_k\ge 0,\forall k \in\mathcal{K}.\nonumber
\end{align}
For notational convenience, let $\mathcal{X}$ denote  the set of $\boldsymbol{\lambda}$ and $\boldsymbol{\mu}$ specified by the constraints in (D2.1). As the strong duality holds between (P2.1) and (D2.1), we can solve (P2.1) by equivalently solving (D2.1). In the following, we first obtain $g(\boldsymbol\lambda,\boldsymbol\mu)$ by solving problem (\ref{g}) under any given $(\boldsymbol{\lambda},\boldsymbol{\mu})\in \mathcal X$, and then solve (D2.1) by finding the optimal $\boldsymbol{\lambda}$ and $\boldsymbol{\mu}$ to minimize $g(\boldsymbol{\lambda},\boldsymbol{\mu})$.
\subsubsection{Obtaining $g(\boldsymbol{\lambda},\boldsymbol{\mu})$ by Solving Problem (\ref{g}) Under Given $(\boldsymbol{\lambda},\boldsymbol{\mu})\in \mathcal{X}$}

First, consider problem (\ref{g}) under any given $(\boldsymbol\lambda,\boldsymbol\mu)\in \mathcal X$. It is evident that problem (\ref{g}) can be decomposed into the following subproblems.
\begin{align}
&~~~~~~~\max_{\substack{R}}~~~~~~~\bigg(1-\sum_{k=1}^K\lambda_k\bigg)R. \label{subL1}\\
&\max_{\substack{\{Q_k(t),\rho_k(t)\},\bold{q}(t)}}~\sum_{k=1}^K\rho_k(t)\varphi_k(\bold{q}(t),Q_k(t),\lambda_k,\mu_k)\nonumber\\
&~~~~~~~~~~~~~~~~~~~~+\rho_0(t)\phi(\bold{q}(t),\{\mu_k\})\label{subL2}\\
&~~~~~~~~\mathrm{s.t.}~~~~~~~Q_k(t) \ge 0,\forall k \in\mathcal{K}\nonumber\\
&~~~~~~~~~~~~~~~~~~~\rho_k(t)\in\{0,1\},\forall k \in\{0\}\cup\mathcal{K},~\sum_{k=0}^K \rho_k(t)= 1,\label{subL2:con}
\end{align}
$\forall t\in\mathcal T$, where
\begin{align}
\varphi_k(\bold{q}(t),Q_k(t),\lambda_k,\mu_k)=&\frac{\lambda_k}{T}{\rm{log_2}}\left(1+\frac{Q_k(t)h_k(\bold{q}(t))}{\sigma^2}\right)\nonumber\\
&-\mu_k Q_k(t),k \in\mathcal{K},\nonumber\\
\phi(\bold{q}(t),\{\mu_k\})=&\sum_{k=1}^K\eta P\mu_kh_k(\bold{q}(t)).\nonumber
\end{align}
Here, problem (\ref{subL2}) consists of an infinite number of subproblems, each corresponding to one time instant $t$.

Note that the optimal value of problem (\ref{subL1}) is always zero as $1 - \sum_{k=1}^K \lambda_k= 0$ (see Lemma 3.1). In this case, the optimal solution $R^*$ to problem (\ref{subL1}) can be chosen as any arbitrary real number. Therefore, we only need to focus on problem (\ref{subL2}). As the subproblems in (\ref{subL2}) are identical for different time instants $t$'s, we can drop the index $t$ for notational convenience, and denote the optimal solution as $\{Q_k^*\},\bold{q}^*$, and $\{\rho^*_k\}$.

As for problem (\ref{subL2}), there are a total of $K+1$ feasible choices for $\{\rho_k\}$ due to the constraints in \eqref{subL2:con}. In the following, we solve problem (\ref{subL2}) by first obtaining the maximum objective value (and the corresponding optimal $\{Q_k\}$ and $\bold{q}$) under each of the $K+1$ feasible $\{\rho_k\}$, and then comparing them to obtain the optimal $\{\rho_k\}$.

First, consider that $\rho_0 = 1$ and $\rho_k = 0, \forall k\in\mathcal K$. In this case, problem (\ref{subL2}) can be re-expressed as
\begin{align}
\max_{\substack{\{Q_k(t)\},\bold{q}}}&~\phi(\bold{q},\{\mu_k\})\nonumber \\
~~\mathrm{s.t.}~&~Q_k \ge 0,\forall k \in\mathcal{K},\label{eqn:new:a}
\end{align}
for which the optimal solution is given as $Q_k = 0, \forall k\in \mathcal K$, and $\bold{q} = \bar{\bold{q}}_\omega^{(\boldsymbol\mu)}, \omega\in\{1,\ldots,\Omega^{(\boldsymbol\mu)}\}$, where
\begin{align}
\{\bar{\bold{q}}^{(\boldsymbol\mu)}_{\omega}\}_{\omega = 1}^{\Omega^{(\boldsymbol\mu)}} &= \arg\max_{\bold{q}}~\phi(\bold{q},\{\mu_k\})\label{qmu}
\end{align}
corresponds to the set of optimal hovering locations for downlink WPT, with $\Omega^{(\boldsymbol\mu)} \ge 1$ denoting the number of optimal solutions to problem \eqref{qmu}. Here, for the non-convex problem (\ref{qmu}), we solve it by using a 2D exhaustive search over the region $[\underline{x},\overline{x}]\times[\underline{y},\overline{y}]$, where $\underline{x}=\min_{k\in\mathcal{K}}x_k,~\overline{x}=\max_{k\in\mathcal{K}}x_k,~\underline{y}=\min_{k\in\mathcal{K}}y_k,~\overline{y}=\max_{k\in\mathcal{K}}y_k$. Note that when the optimal solution to problem \eqref{qmu} is non-unique (or $\Omega^{(\boldsymbol{\mu})} > 1$), we can arbitrarily choose any one of $\bar{\bold{q}}^{(\boldsymbol\mu)}_\omega$'s for obtaining the dual function $g(\boldsymbol{\lambda},\boldsymbol{\mu})$. Accordingly, the optimal value of problem (\ref{subL2}) is given as $\phi(\bar{\bold{q}}^{(\boldsymbol\mu)}_\omega,\{\mu_k\})$.


Next, consider that $\rho_k= 1$ for any one $k\in\mathcal K$ and $\rho_j = 0, \forall j\in \{0\}\cup\mathcal K,j\neq k$. In this case, problem (\ref{subL2}) can be re-expressed as
\begin{align}
&\max_{\substack{Q_k,\bold{q}}}~\varphi_k(\bold{q},Q_k,\lambda_k,\mu_k)\nonumber\\
&~~\mathrm{s.t.}~~Q_j \ge 0,\forall j \in\mathcal{K}.\label{eqn:new:b}
\end{align}
Note that the objective function of problem \eqref{eqn:new:b} is concave with respect to $Q_k$, and therefore, problem \eqref{eqn:new:b} is convex. By checking the Karush-Kuhn-Tucker (KKT) conditions, we have the optimal solution as $\bold{q}=\bold{w}_k$, $Q_k = Q^{(\lambda_k,\mu_k)} \triangleq \left(\frac{\lambda_k}{T\mu_k\ln 2}-\frac{H^2}{\gamma}\right)^+$, and $Q_j = 0, \forall j\in\mathcal K,j\neq k$, where $(x)^+\triangleq\max(x,0)$. Therefore, the corresponding optimal value of problem (\ref{subL2}) is $\varphi_k(\bold{w}_k,Q^{(\lambda_k,\mu_k)},\lambda_k,\mu_k)$.

By comparing the $K+1$ optimal values, i.e., $\phi(\{\bar{\bold{q}}^{(\boldsymbol\mu)}_\omega\},\{\mu_k\})$ and $\varphi_k(\bold{w}_k,Q^{(\lambda_k,\mu_k)},\lambda_k,\mu_k)$, $\forall k\in\mathcal K$, we have the following proposition, for which the proof is straightforward and thus omitted.
\begin{proposition}\label{proposition:3.1}
The optimal solution to problem (\ref{subL2}) is obtained by considering following two cases.
\begin{itemize}
\item If $\phi(\{\bar{\bold{q}}^{(\boldsymbol\mu)}_\omega\},\{\mu_k\}) \ge \varphi_k(\bold{w}_k,Q^{(\lambda_k,\mu_k)},\lambda_k,\mu_k), \forall k\in\mathcal K$, then the UAV operates in the downlink WPT mode, i.e., 
\begin{align}
&\rho_0^*=1, ~\rho_k^*= 0,~Q_k^*=0, \forall k\in\mathcal K,\nonumber\\
&\bold{q}^* \in \{\bar{\bold{q}}^{(\boldsymbol\mu)}_1,\ldots,\bar{\bold{q}}^{(\boldsymbol\mu)}_{\Omega^{(\boldsymbol\mu)}}\},\label{omega>k}
\end{align}
where $\bold{q}^*$ is generally non-unique when $\Omega^{(\boldsymbol\mu)} >1$.
\item Otherwise, we denote
$k^* = \arg\max_{k\in\mathcal K}~ \varphi_k(\bold{w}_k,Q^{(\lambda_k,\mu_k)},\lambda_k,\mu_k)$.
Then the UAV operates in the uplink WIT mode for user $k^*$, i.e.,
\begin{align}
&\rho_0^*=0, ~\rho_{k^*}^*= 1,~\rho_{j}^*= 0,\forall j\in\mathcal K, j\neq k^*,\nonumber\\
&Q_{k^*}^*=Q^{(\lambda_{k^*},\mu_{k^*})},~Q_j^*=0,\forall j\in\mathcal K,j\neq k^*, \nonumber\\
&\bold{q}^* = {\bold{w}}_{k^*}.\label{k>omega}
\end{align}
%
%
\end{itemize}
\end{proposition}

Note that if any two of the $K+1$ optimal values (i.e., $\varphi_k(\bold{w}_k,Q^{(\lambda_k,\mu_k)},\lambda_k,\mu_k), \forall k\in\mathcal K$ and $\phi(\{\bar{\bold{q}}^{(\boldsymbol\mu)}_\omega\},\{\mu_k\})$) are equal, then the corresponding solutions in (\ref{omega>k}) and (\ref{k>omega}) are both optimal for problem (\ref{subL2}). Based on Proposition 3.1, problem (\ref{subL2}) is solved, and thus the function $g(\boldsymbol{\lambda},\boldsymbol{\mu})$ is obtained.

\subsubsection{Finding Optimal $\boldsymbol\lambda$ and $\boldsymbol\mu$ to Solve (D2.1)}
Next, we search over ($\boldsymbol{\lambda},\boldsymbol{\mu}$) to minimize $g(\boldsymbol{\lambda},\boldsymbol{\mu})$ for solving (D2.1). Since the dual problem (D2.1) is always convex but in general non-differentiable, we can use subgradient based methods, such as the ellipsoid method\cite{Boyd2004}, to obtain the optimal $\boldsymbol{\lambda}$ and $\boldsymbol{\mu}$, denoted by $\boldsymbol{\lambda}^{\text{opt}}$ and $\boldsymbol{\mu}^{\text{opt}}$. Note that for the objective function $g(\boldsymbol{\lambda},\boldsymbol{\mu})$ in (D2.1), the subgradient with respect to ($\boldsymbol{\lambda},\boldsymbol{\mu}$) is
\begin{align}
&\bigg[r_1(\rho_1^*,\bold{q}^*,Q_1^*),\ldots,r_K(\rho_K^*,\bold{q}^*,Q_K^*),\nonumber\\
&T E_1(\rho_0^*,\bold{q}^*,P^*)-T\rho_1^*Q_1^*,\ldots,T E_K(\rho_0^*,\bold{q}^*,P^*)-T\rho_K^*Q_K^*\bigg],\nonumber
\end{align}
where $R^*=0$ is chosen for simplicity. 
\subsubsection{Constructing Optimal Primal Solution to (P2.1)}

With $\boldsymbol{\lambda}^{\text{opt}}$ and $\boldsymbol{\mu}^{\text{opt}}$ at hand, it remains to construct the optimal primal solution to (P2.1), denoted by $\{\rho_k^{\text{opt}}(t),Q^{\text{opt}}_k(t),\bold{q}^{\text{opt}}(t)\}$ and $R^{\text{opt}}$. Before proceeding, we have the following proposition.
\begin{proposition}\label{proposition:3.2}
It must hold that
$\phi(\{\bar{\bold{q}}_\omega^{(\boldsymbol\mu^{\text{opt}})}\},\{\mu_k^{\text{opt}}\}) = \varphi_k(\bold{w}_k,Q^{(\lambda_k^{\text{opt}},\mu_k^{\text{opt}})},\lambda_k^{\text{opt}},\mu_k^{\text{opt}}), \forall k\in\mathcal K, \omega\in\{1,\dots,\Omega^{(\boldsymbol\mu^{\text{opt}})}\}$ at the optimal $\boldsymbol{\lambda}^{\text{opt}}$ and $\boldsymbol{\mu}^{\text{opt}}$.
\end{proposition}
\begin{IEEEproof}
See Appendix \ref{Appendix:2}.
\end{IEEEproof}

By combining Propositions \ref{proposition:3.1} and \ref{proposition:3.2}, it follows that under the optimal dual solution $\boldsymbol{\lambda}^{\text{opt}}$ and $\boldsymbol{\mu}^{\text{opt}}$ to (D2.1), problem (\ref{subL2}) has a total number of $\Omega^{(\boldsymbol{\mu}^{\text{opt}})}+K$ optimal solutions. Among them, the $\Omega^{(\boldsymbol{\mu}^{\text{opt}})}$ optimal solutions are given in \eqref{omega>k} for downlink WPT, and the other $K$ solutions are given in \eqref{k>omega} for uplink WIT (each for one user $k$). In this case, we need to time-share among these optimal solutions to construct the optimal primal solution to (P2.1).

More specifically, notice that the $\Omega^{(\boldsymbol{\mu}^{\text{opt}})}$ solutions in \eqref{omega>k} correspond to $\Omega^{(\boldsymbol{\mu}^{\text{opt}})}$ hovering locations $\bar{\bold{q}}^{(\boldsymbol\mu^{\text{opt}})}_1,\ldots,\bar{\bold{q}}^{(\boldsymbol\mu^{\text{opt}})}_{\Omega^{(\boldsymbol\mu^{\text{opt}})}}$ for downlink WPT, at which only the UAV transmits at constant power $P$ with $\rho_0^*=1$; on the other hand, the $k$-th solution in \eqref{k>omega}, $k\in\mathcal K$, corresponds to that the UAV hovers above user $k$ at location $\bold{w}_k$ for uplink WIT, at which user $k$ transmits with $Q_k^*=Q^{(\lambda_k^{\text{opt}},\mu_k^{\text{opt}})}$ and $\rho_k^*=1$. With time-sharing, let $\tau_\omega$ and $\varsigma_k$ denote the hovering durations at the location $\bar{\bold{q}}^{(\boldsymbol\mu^{\text{opt}})}_\omega, \omega \in\{1,\ldots,\Omega^{(\boldsymbol\mu^{\text{opt}})}\}$, and $\bold{w}_k, k\in\{1,\ldots,K\}$, respectively. In this case, we solve the following uplink common throughput maximization problem to obtain the optimal hovering durations $\tau_\omega$'s and $\varsigma_k$'s for time sharing.
\begin{align}
\text{(P2.2)}:&~\max_{\substack{\{\varsigma_k \ge 0,\tau_\omega \ge 0\},R}} R\nonumber\\
\mathrm{s.t.}~&\frac{\varsigma_k}{T}{\rm{log_2}}\left(1+\frac{Q^{(\lambda_k^{\text{opt}},\mu_k^{\text{opt}})}h_k(\bold{w}_k)}{\sigma^2}\right)\ge R,\forall k\in\mathcal{K}\\
~&\varsigma_kQ^{(\lambda_k^{\text{opt}},\mu_k^{\text{opt}})} \le \sum_{\omega=1}^{\Omega^{(\boldsymbol\mu^{\text{opt}})}} \tau_\omega\eta Ph_k(\bar{\bold{q}}^{(\boldsymbol\mu^{\text{opt}})}_{\omega}),\forall k\in\mathcal{K}\\
~&\sum_{k=1}^{K} \varsigma_k+\sum_{\omega=1}^{\Omega} \tau_\omega= T.
\end{align}
Note that problem (P2.2) is a linear program, which can be solved by standard convex optimization techniques in\cite{Boyd2004}. The optimal solution to (P2.2) is denoted as $\hat \varsigma_k$, $\hat\tau_\omega$ and $\hat R$. Accordingly, we can divide the whole period $\mathcal T$ into $\Omega^{(\boldsymbol\mu^{\text{opt}})} + K$ sub-periods, where the first $\Omega^{(\boldsymbol\mu^{\text{opt}})}$ sub-periods, denoted by $\mathcal T_\omega = (\sum_{j=1}^{\omega-1} \hat\tau_j,\sum_{j=1}^{\omega} \hat\tau_j], \omega\in\{1,\ldots,\Omega^{(\boldsymbol\mu^{\text{opt}})}\}$, are for downlink WPT, and the next $K$ sub-periods, denoted by $\mathcal T_{\Omega^{(\boldsymbol\mu^{\text{opt}})}+k}=(\sum_{\omega=1}^{\Omega^{(\boldsymbol\mu^{\text{opt}})}} \hat\tau_\omega + \sum_{j=1}^{k-1}\hat\varsigma_j ,\sum_{\omega=1}^{\Omega^{(\boldsymbol\mu^{\text{opt}})}} \hat\tau_\omega + \sum_{j=1}^{k}\hat\varsigma_j]$, $k\in\mathcal K$, are for uplink WIT of the $K$ users. As a result, we have the following proposition, for which the proof is omitted for brevity.
\begin{proposition}\label{proposition:3.3}
The optimal solution to (P2.1) (and thus (P2)) is given as follows. During sub-period $\omega \in \{1,\ldots,\Omega^{(\boldsymbol\mu^{\text{opt}})}\}$, the UAV hovers at the location $\bar{\bold{q}}^{(\boldsymbol\mu^{\text{opt}})}_\omega$ for downlink WPT, i.e.,
\begin{align}
\bold{q}^{\text{opt}}(t) &= \bar{\bold{q}}^{(\boldsymbol\mu^{\text{opt}})}_\omega,
\rho_0^{\text{opt}}(t) = 1, ~\rho_k^{\text{opt}}(t) = 0,~Q^{\text{opt}}_k(t) = 0, \forall k\in\mathcal K,
\end{align}
$\forall t\in \mathcal T_\omega, \omega \in \{1,\ldots,\Omega^{(\boldsymbol\mu^{\text{opt}})}\}$. During sub-period $\Omega^{(\boldsymbol\mu^{\text{opt}})} + k$, $k\in\mathcal K$, the UAV hovers above user $k$ at $\bold{w}_k$, and user $k$ sends information to the UAV in the uplink, i.e.,
\begin{align}
&\bold{q}^{\text{opt}}(t) = {\bold{w}}_k,~\rho_k^{\text{opt}}(t) = 1,~Q^{\text{opt}}_k(t) = Q^{(\lambda_k^{\text{opt}},\mu_k^{\text{opt}})},\nonumber\\ 
&\rho_0^{\text{opt}}(t) = 0,~\rho_j^{\text{opt}}(t) = 0,~Q^{\text{opt}}_j(t)= 0,\forall j\in\mathcal K,j\neq k,
\end{align}
$\forall t\in \mathcal T_{\Omega^{(\boldsymbol\mu^{\text{opt}})} + k}, k \in \mathcal K$. The optimal uplink common throughput is given as $R^{\text{opt}} = \hat{R}$ (with $\hat{R}$ denoting the optimal solution obtained for (P2.2)).
\end{proposition}

In summary, we present the overall algorithm for solving (P2) as Algorithm 1 in Table \ref{algorithm:1}, and we refer to such a solution as the {\it{multi-location-hovering}} solution. Notice that Algorithm 1 is guaranteed to converge to the globally optimal solution to problem (P2).
\begin{table}
\begin{center}
\caption{Algorithm 1 for Solving Problem (P2)}  \vspace{0.1cm}
\hrule \vspace{0.3cm}
\begin{itemize}
\item[a)] {\bf Initialization:} Given an ellipsoid $\boldsymbol{\varepsilon}((\boldsymbol{\lambda},\boldsymbol{\mu}),\bf{A})$ containing ($\boldsymbol{\lambda}^{\text{opt}},\boldsymbol{\mu}^{\text{opt}}$), where ($\boldsymbol{\lambda},\boldsymbol{\mu}$) is the center point of $\boldsymbol{\varepsilon}((\boldsymbol{\lambda},\boldsymbol{\mu}),\bf{A})$ and the positive definite matrix $\bf{A}$ characterizes the size of $\boldsymbol{\varepsilon}((\boldsymbol{\lambda},\boldsymbol{\mu}),\bf{A})$.
\item[b)] {\bf Repeat:}
    \begin{itemize}
    \item[1)] Obtain $\{\bar{\bold{q}}^{(\boldsymbol\mu)}_\omega\}$ to maximize $\phi(\bold{q},\{\mu_k\})$ in problem (\ref{qmu}) via a 2D exhaustive search over the region $[\underline{x},\overline{x}]\times[\underline{y},\overline{y}]$;
    \item[2)] Obtain $g(\boldsymbol{\lambda},\boldsymbol{\mu})$ under given ($\boldsymbol{\lambda},\boldsymbol{\mu}$) by using Proposition \ref{proposition:3.1};
    \item[3)] Compute the subgradients of $g(\boldsymbol{\lambda},\boldsymbol{\mu})$, and then update $\boldsymbol{\lambda}$ and $\boldsymbol{\mu}$ by using the ellipsoid method \cite{Boyd2004}.
    \end{itemize}
\item[c)] {\bf Until} $\boldsymbol{\lambda}$ and $\boldsymbol{\mu}$ converge with a prescribed accuracy.
\item[d)] {\bf{Set}} ($\boldsymbol{\lambda}^{\text{opt}},\boldsymbol{\mu}^{\text{opt}})$$\leftarrow$
    ($\boldsymbol{\lambda},\boldsymbol{\mu}$).
\item[e)] Obtain the optimal solution to problem (P2.1) or (P2) based on Proposition \ref{proposition:3.3}.
\end{itemize}
\vspace{0.1cm} \hrule
\label{algorithm:1}
\end{center}
\end{table}

\begin{remark}\label{Remark:3.1}
It is worth noting that similar multi-location-hovering solutions have been proposed in the UAV-enabled multiuser WPT system \cite{JieXuWPT} and the UAV-enabled multiuser communication system with TDMA transmission \cite{Qingqingwu}, when the UAV's flight period becomes sufficiently long. In \cite{JieXuWPT}, the UAV successively hovers above a given set of locations to maximize all users' minimum received energy; while in \cite{Qingqingwu}, the UAV successively hovers above each user to maximize the minimum throughput of all users. As a matter of fact, our derived multi-location-hovering solution to problem (P2) in Proposition \ref{proposition:3.3} unifies the results in \cite{JieXuWPT} and \cite{Qingqingwu}, which consists of two sets of hovering locations: $\Omega^{(\boldsymbol{\mu}^{\text{opt}})}$ ones for WPT and the other $K$ ones for WIT. Nevertheless, note that the $\Omega^{(\boldsymbol{\mu}^{\text{opt}})}$ hovering locations to problem (P2) for WPT are generally different from those in \cite{JieXuWPT}, as they are designed based on different objective functions (max-min communication throughput versus max-min harvested energy).
\end{remark}

\begin{figure}
\begin{minipage}[t]{0.48\textwidth}
\centering
    \includegraphics[width=7cm]{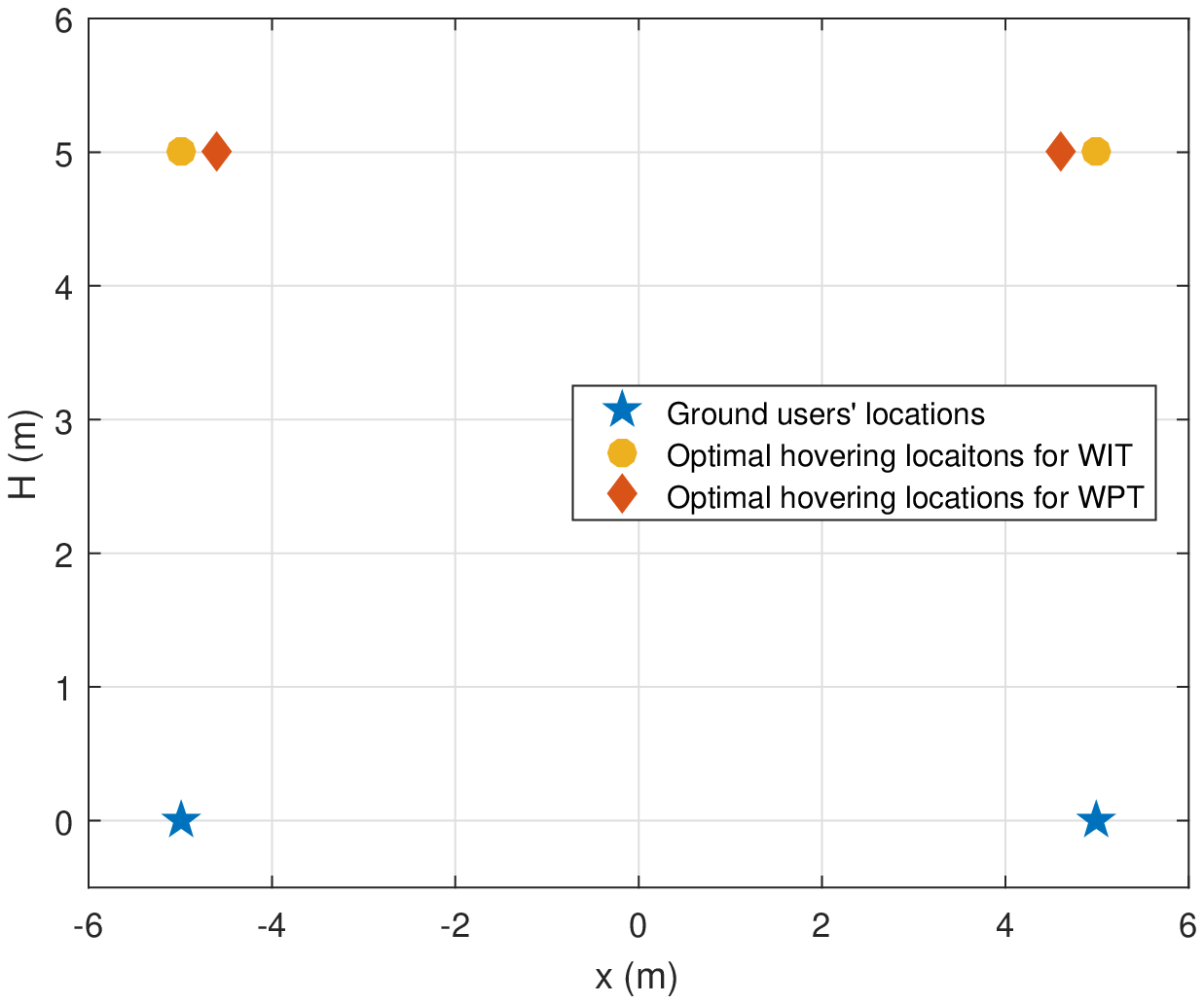}
\caption{The optimal hovering locations for WPT and WIT in a two-user WPCN, where $D=10$~m and $H=5$~m with $D>2H/\sqrt{3}$.} \label{fig:twouser2}
\end{minipage}
\begin{minipage}[t]{0.48\textwidth}
\centering
    \includegraphics[width=7cm]{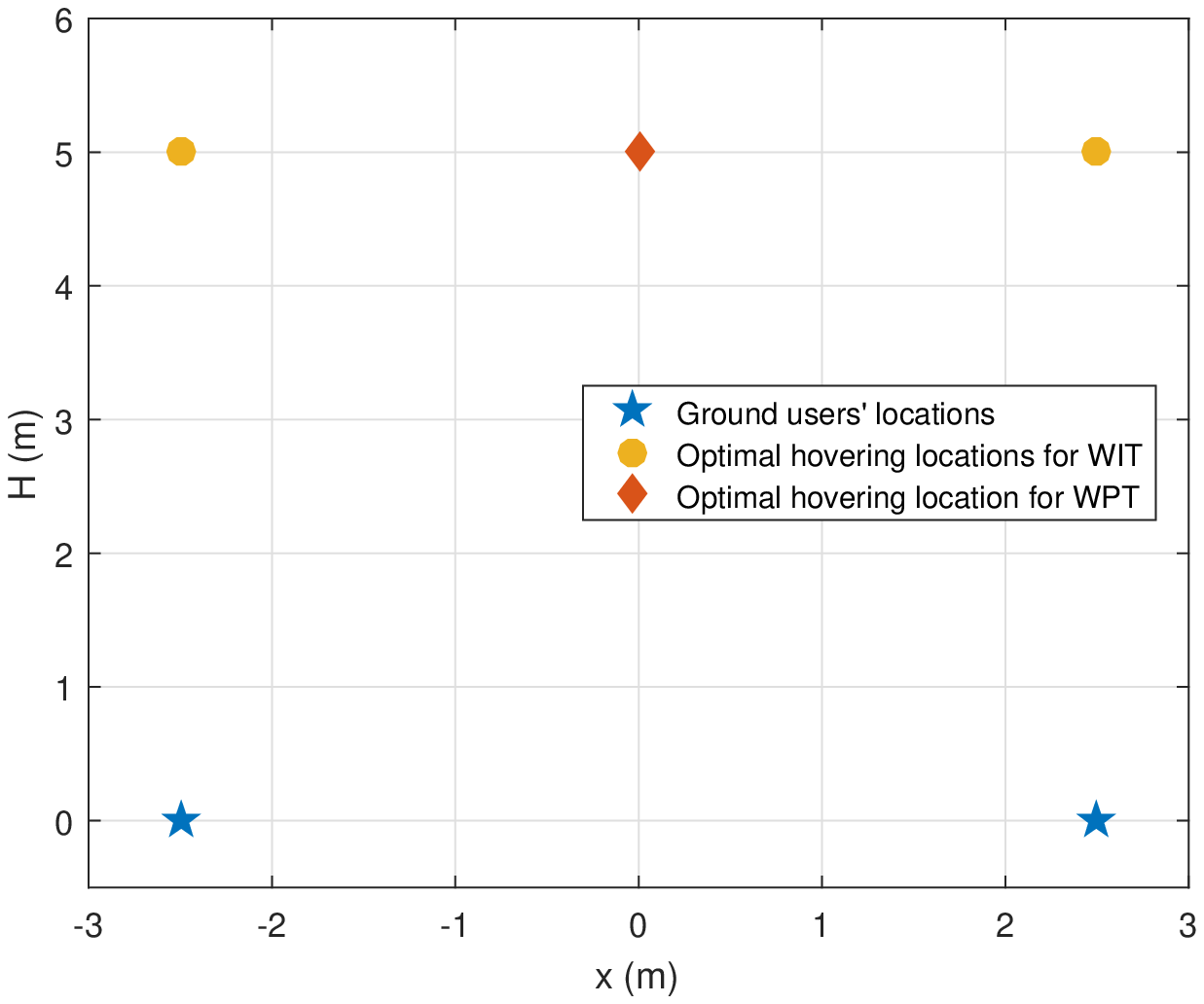}
\caption{The optimal hovering locations for WPT and WIT in a two-user WPCN, where $D=5$~m and $H=5$~m with $D\le2H/\sqrt{3}$.} \label{fig:twouser1}
\end{minipage}
\end{figure}

\begin{remark}\label{twousers}
To gain more insights, it is interesting to consider problem (P2) in the special case of $K=2$ users. Without loss of generality, we assume that the two users are located at $(-D/2,0,0)$ and $(D/2,0,0)$ with $x_1=-D/2$, $x_2=D/2$, and $y_1=y_2=0$, where $D$ denotes the distance between the two users. Due to this symmetric setup, it can be shown that the optimal $\Omega^{(\boldsymbol{\mu}^{\text{opt}})}$ hovering locations for WPT to problem (P2) are actually identical to the optimal hovering locations to maximize the two users' minimum harvested energy in the UAV-enabled WPT system \cite{JieXuWPT,Jiewpt}. It follows from \cite{JieXuWPT,Jiewpt} that the optimal hovering locations for downlink WPT are critically dependent on the UAV's flying altitude $H$ and the users' distance $D$. In particular, when $D>2H/\sqrt{3}$, there are $\Omega^{(\boldsymbol{\mu}^{\text{opt}})}$= 2 hovering locations at $(-\epsilon,0,H)$ and $(\epsilon,0,H)$ for WPT, where $\epsilon\triangleq \sqrt{-(D^2/4+H^2)+\sqrt{D^4/4+H^2D^2}}$ with $\lim_{D\to\infty} \epsilon = D/2$; while when $D\le2H/\sqrt{3}$ there is only $\Omega^{(\boldsymbol{\mu}^{\text{opt}})}$ = 1 hovering location $(0,0,H)$ right above the middle point of two users. By combining the optimal hovering locations for WPT and those for WIT, it is evident that when $D>2H/\sqrt{3}$, the UAV needs to hover above four locations for efficient WPCN, as shown by the example in  Fig. \ref{fig:twouser2}; while when $D\le2H/\sqrt{3}$, the UAV needs to hover above three locations, as shown by the example in Fig. \ref{fig:twouser1}.
%
%
%

%
\end{remark}

\section{Proposed Solution to Problem (P1) with Successive Hover-and-Fly Trajectory}\label{sec:P1}
This section considers problem (P1) with the UAV's maximum speed constraint considered. First, we present a successive hover-and-fly trajectory motivated by the multi-location-hovering solution to the relaxed problem (P2), in which the UAV sequentially visits the $\Omega^{(\boldsymbol{\mu}^{\text{opt}})} + K$ hovering locations for efficient WPT and WIT, respectively. Next, under such a flying trajectory, we design the duration at each hovering location and the transmission resource allocation for (P1) by discretizing the time period. Finally, we discuss the case when the UAV's flight duration $T$ is too small to visit all these hovering locations, for which the trajectory and transmission resource allocations are redesigned.
\subsection{Successive Hover-and-Fly UAV Trajectory}
In the proposed successive hover-and-fly trajectory design, the UAV sequentially visits the ${\Omega^{(\boldsymbol\mu^{\text{opt}})}}+K$ optimal hovering locations that are obtained for (P2), i.e., $\bar{\bold{q}}^{(\boldsymbol\mu^{\text{opt}})}_1,\ldots,\bar{\bold{q}}^{(\boldsymbol\mu^{\text{opt}})}_{\Omega^{(\boldsymbol\mu^{\text{opt}})}}$ for downlink WPT, and $\bold{w}_1,\ldots,\bold{w}_K$ for uplink WIT. For notational convenience, we denote the ${\Omega^{(\boldsymbol\mu^{\text{opt}})}}+K$ hovering locations as $\{\bold{q}_k^{\text{opt}}\}$, where $\bold{q}_k^{\text{opt}} = \bar{\bold{q}}^{(\boldsymbol\mu^{\text{opt}})}_{k},  k\in\{1,\ldots,\Omega^{(\boldsymbol\mu^{\text{opt}})}\}$, and $\bold{q}_{\Omega^{(\boldsymbol\mu^{\text{opt}})}+k}^{\text{opt}} = \bold{w}_{k}, k\in\mathcal K$. In order to maximize the time for efficient WPT and WIT, the UAV flies among these hovering locations by using the maximum speed $V_{\max}$, and the UAV aims to minimize the flying time by equivalently minimizing the traveling path among the ${\Omega^{(\boldsymbol\mu^{\text{opt}})}}+K$ locations.

Towards this end, we define a set of binary variables $\{f_{j,k}\}, \forall j,k\in\{1,\dots,{\Omega^{(\boldsymbol\mu^{\text{opt}})}}+K\},j\neq k$, where $f_{j,k}=1$ or $f_{j,k}=0$ indicates that the UAV flies or does not fly from the $j$-th hovering location $\bold{q}_j^{\text{opt}}$ to the $k$-th hovering location $\bold{q}_k^{\text{opt}}$. Hence, the traveling path minimization problem becomes determining $\{f_{j,k}\}$ to minimize $\sum_{j=1}^{{\Omega^{(\boldsymbol\mu^{\text{opt}})}}+K}\sum_{k=1,k\neq j}^{{\Omega^{(\boldsymbol\mu^{\text{opt}})}}+K}f_{j,k}d_{j,k}$, provided that each of the $\Omega^{(\boldsymbol\mu^{\text{opt}})}+K$ locations is visited once, where $d_{j,k} = {\|\bold{q}_{j}^{\text{opt}} - \bold{q}_{k}^{\text{opt}}\|}$ denotes the distance between $\bold{q}_{j}^{\text{opt}} $ and $\bold{q}_{k}^{\text{opt}}$. Note that as shown in \cite{JieXuWPT}, the flying distance minimization is similar to the well-established traveling salesman problem (TSP), with the following differences. The standard TSP requires the salesman (or the UAV in this paper) to return to the origin city (the initial hovering location) after visiting all the other cities (or hovering locations here), while the flying distance minimization problem of our interest does not have such a requirement since the initial and final hovering locations can be optimized. As shown in \cite{TSP}, we can transform our traveling distance minimization problem to the standard TSP as follows. First, we add a dummy hovering location, namely the $(\Omega^{(\boldsymbol\mu^{\text{opt}})}+K+1)$-th hovering location, whose distances to all the existing $\Omega^{(\boldsymbol\mu^{\text{opt}})}+K$ hovering locations are 0, i.e., $d_{\Omega^{(\boldsymbol\mu^{\text{opt}})}+K+1,k}=d_{k,\Omega^{(\boldsymbol\mu^{\text{opt}})}+K+1}=0$, $\forall k\in\{1,\dots,\Omega^{(\boldsymbol\mu^{\text{opt}})}+K\}$. Note that this dummy hovering location is a virtual node that does not exist physically. Then, we obtain the desirable traveling path by solving the standard TSP problem for the $\Omega^{(\boldsymbol\mu^{\text{opt}})}+K+1$ hovering locations, and then removing the two edges associated with the dummy location.  As a result, we use the permutation $\kappa(\cdot)$ over the set $\{1,\dots,{\Omega^{(\boldsymbol\mu^{\text{opt}})}}+K\}$ to denote the obtained traveling path, where the $\kappa(1)$-th hovering location is first visited, followed by the $\kappa(2)$-th, the $\kappa(3)$-th, etc., until the $\kappa(\Omega^{(\boldsymbol\mu^{\text{opt}})}+K)$-th hovering location at last. We denote the traveling distance and traveling duration from the $\kappa(i)$-th hovering location $\bold{q}_{\kappa(i)}^{\text{opt}}$ to the $\kappa(i+1)$-th hovering location $\bold{q}_{\kappa(i+1)}^{\text{opt}}$ as $d_{\kappa(i),\kappa(i+1)}$ and $T_{\text{fly},i}=d_{\kappa(i),\kappa(i+1)}/V_{\max}$, respectively, $i\in\{1,...,{\Omega^{(\boldsymbol\mu^{\text{opt}})}}+K-1\}$. Hence, the total traveling distance and duration are given as $D_{\text{fly}}=\sum_{i=1}^{{\Omega^{(\boldsymbol\mu^{\text{opt}})}}+K-1}d_{\kappa(i),\kappa(i+1)}$, and $T_{\text{fly}}=D_{\text{fly}}/V_{\max}$, respectively. We denote the obtained flying trajectory as $\{\tilde{\bold{q}}(t)\}_{t=0}^{T_{\text{fly}}}$.

Note that in practice, the UAV's flight duration $T$ may not be exactly equal to the traveling distance $T_{\text{fly}}$. When $T > T_{\text{fly}}$, we further need to determine the hovering durations at the $\Omega^{(\boldsymbol\mu^{\text{opt}})} + K$ locations. When $T<T_{\text{fly}}$, the UAV does not have sufficient time to visit all the $\Omega^{(\boldsymbol\mu^{\text{opt}})} + K$ hovering locations, and thus we need to redesign the UAV trajectory in order to satisfy the duration requirements. We present the complete successive hover-and-fly trajectory for the above two cases in Sections \ref{TgeTT} and \ref{T<T}, respectively, together with the transmission resource allocations.
\subsection{Hovering Durations and Transmission Resource Allocation Optimization When $T> T_{\text{fly}}$}\label{TgeTT}
Let $\tau_\omega$ denote the time duration for the UAV to hover at the location $\bold{q}_{\omega}^{\text{opt}} = \bar{\bold{q}}^{(\boldsymbol\mu^{\text{opt}})}_\omega$ for WPT, $\omega\in\{1,\ldots,{\Omega^{(\boldsymbol\mu^{\text{opt}})}}\}$, and $\varsigma_k$ denote the  the time duration for the UAV to hover above user $k$ at $\bold{q}_{\Omega^{(\boldsymbol\mu^{\text{opt}})}+k}^{\text{opt}} = \bold{w}_{k}$ for WIT, $k\in\{1,\ldots,K\}$. Furthermore, we define
\begin{align*}
\upsilon_{k} \triangleq
\left\{
\begin{array}{ll}
\tau_{\kappa(k)}, & {\text{if}}~ \kappa(k)\in\{1,\ldots,\Omega^{(\boldsymbol\mu^{\text{opt}})}\} \\
\varsigma_{\kappa(k) - \Omega^{(\boldsymbol\mu^{\text{opt}})}},& {\text{if}}~ \kappa(k)\in \{\Omega^{(\boldsymbol\mu^{\text{opt}})} + 1,\ldots,\Omega^{(\boldsymbol\mu^{\text{opt}})} + K\}.
\end{array}
\right.
\end{align*}
Accordingly, we can divide the time duration into $2({\Omega^{(\boldsymbol\mu^{\text{opt}})}} + K)-1$ sub-periods, denoted by $[\hat{\mathcal T}_1,\hat{\mathcal T}_2,\dots,\hat{\mathcal T}_{2({\Omega^{(\boldsymbol\mu^{\text{opt}})}} + K)-1}]$, which are defined as $\hat{\mathcal{T}}_{2k - 1} \triangleq
\bigg(\sum_{i=1}^{k-1}(\upsilon_{\kappa(i)}+T_{\text{fly},i}), \sum_{i=1}^{k-1}(\upsilon_{\kappa(i)}+T_{\text{fly},i})+\upsilon_{\kappa(k)}\bigg]$ for odd sub-periods with $k\in\{1,\ldots, \Omega^{(\boldsymbol\mu^{\text{opt}})} + K\}$, and $\hat{\mathcal{T}}_{2k} \triangleq \bigg(\sum_{i=1}^{k-1}(\upsilon_{\kappa(i)}+T_{\text{fly},i})+\upsilon_{\kappa(k)}, \sum_{i=1}^{k}(\upsilon_{\kappa(i)}+T_{\text{fly},i})\bigg]$ for even sub-periods with $k\in\{1,\ldots, \Omega^{(\boldsymbol\mu^{\text{opt}})} + K -1\}$. Therefore, within each odd sub-period $2k - 1$, the UAV should hover at the $\kappa(k)$-th hovering location $(x_{\kappa(k)}, y_{\kappa(k)},H)$, i.e.,
\begin{align}
\bold{q}(t) = \bold{q}_{\kappa(k)}^{\text{opt}}, \forall t\in \hat{\mathcal{T}}_{2k - 1}, ~k\in\{1,\ldots, \Omega^{(\boldsymbol\mu^{\text{opt}})} + K\}. \label{eqn:trajectory:1}
\end{align}
During each even sub-period $2k$, the UAV should fly from the $\kappa(k)$-th hovering location to the $\kappa(k+1)$-th hovering location with the UAV's maximum speed $V_{\max}$, whose time-varying location is
\begin{align}
\bold{q}(t) =  \tilde{\bold{q}}\bigg(t-\sum_{i=1}^{k} \upsilon_{\kappa(i)}\bigg),\label{eqn:trajectory:2}
\end{align}
$\forall t\in \hat{\mathcal{T}}_{2k},k\in\{1,\ldots, \Omega^{(\boldsymbol\mu^{\text{opt}})} + K-1\}$. Therefore, the  successive hover-and-fly trajectory is finally obtained, in which the hovering durations $\{ \upsilon_{\kappa(k)}\}$, or equivalently, $\{\tau_\omega\}$ and $\{\varsigma_k\}$, are optimization variables that will be determined next.

%
%
%
%
%

Under the obtained successive hover-and-fly trajectory, we maximize the uplink common throughput of all users by optimizing the transmission mode $\{\rho_k(t)\}$ and the power allocation $\{Q_k(t)\}$, jointly with the hovering durations $\{\tau_\omega\}$ and $\{\varsigma_k\}$. Towards this end, we first discuss the transmission policy of the UAV-enabled WPCN during each sub-period.

First, consider the $(2k - 1)$-th sub-period $\hat{\mathcal{T}}_{2k - 1}, k\in\{1,\ldots,\Omega^{(\boldsymbol\mu^{\text{opt}})} + K\}$, in which the UAV hovers above $\bold{q}_{\kappa(k)}^{\text{opt}}$ as in \eqref{eqn:trajectory:1}. The transmission policy in this sub-period is based on the optimal solution to (P2) in Section \ref{sec:optimal}. In particular, if $1\le \kappa(k)\le \Omega^{(\boldsymbol\mu^{\text{opt}})}$, we denote $\omega = \kappa(k)$ for convenience. In this case, the UAV hovers at $\bar{\bold{q}}^{(\boldsymbol\mu^{\text{opt}})}_\omega$ with duration $\tau_\omega$, and the UAV works in the downlink WPT mode during this sub-period, by employing transmit power $P$. On the other hand, if $\Omega^{(\boldsymbol\mu^{\text{opt}})}+1\le \kappa(k)\le \Omega^{(\boldsymbol\mu^{\text{opt}})}+K$, we denote $\hat{k} = \kappa(k) - \Omega^{(\boldsymbol\mu^{\text{opt}})}$. In this case, the UAV hovers above user $\hat{k}$ at $\bold{w}_{\hat{k}}$ with duration $\varsigma_{\hat{k}}$, and user $\hat{k}$ works in the uplink WIT mode to transmit information to the UAV during the whole sub-period, by employing a certain transmit power, denoted by $Q^{\text{hover}}_{\hat{k}}$. By combining the $\Omega^{(\boldsymbol\mu^{\text{opt}})} + K$ odd sub-periods, the uplink throughput and the harvested energy at each user $k$ are respectively given as
\begin{align}
\widetilde{R_k}(\varsigma_k,Q^{\text{hover}}_k)&=\frac{\varsigma_k}{T}\log_2\left(1+\frac{h_k(\bold{w}_k)Q^{\text{hover}}_k}{\sigma^2}\right) \label{eqn:33}\\
\widetilde{E_k}(\{\tau_\omega\})&=\sum_{\omega=1}^{\Omega^{(\boldsymbol\mu^{\text{opt}})}} \eta P h_k(\bar{\bold{q}}^{(\boldsymbol\mu^{\text{opt}})}_\omega)  \tau_\omega . \label{eqn:34}
\end{align}

Next, consider even sub-periods $\hat{\mathcal{T}}_{2k}, \forall k\in\{1,\ldots,\Omega^{(\boldsymbol\mu^{\text{opt}})} + K-1\}$, with total duration $T_{\text{fly}}$. We discretize these sub-periods into $N$ slots each with duration $\delta = T_{\text{fly}}/N$. For each slot $n\in\mathcal N\triangleq \{1,\ldots,N\}$, the location of the UAV is assumed to be constant and denoted as $\bold{q}^{\text{fly}}[n] = \tilde{\bold{q}}(n\delta)$. Furthermore, in order to handle the binary transmission mode indicator $\{\rho_k(t)\}$, we consider that the $(K+1)$ transmission modes are time shared within one slot, by dividing each slot $n$ into $(K+1)$ sub-slots without loss of generality. In the first sub-slot with duration $\tau^{\text{fly}}_0[n] \ge 0$, the UAV works in the downlink WPT mode with transmit power $P$, and in sub-slot $k+1$ with duration $\tau^{\text{fly}}_k[n] \ge 0$, user $k$ works in the uplink WIT mode by using transmit power $Q_k^{\text{fly}}[n]$, $k\in\mathcal K$. Here, it follows that $\sum_{k=0}^K \tau_k^{\text{fly}}[n]= \delta, \forall n \in\mathcal{N}$.
By combining the $N$ slots over the even sub-periods, the uplink throughput and the harvested energy at each user $k$ are respectively given by

\begin{footnotesize}
\begin{align}
R^{\text{fly}}_k(\{\tau_k^{\text{fly}}[n],{Q}^{\text{fly}}_k[n]\})= & \frac{1}{T}\sum_{n=1}^N\tau_{k}^{\text{fly}}[n]{\rm{log_2}}\left(1+\frac{h_k(\bold{q}^{\text{fly}}[n]){Q}^{\text{fly}}_k[n]}{\sigma^2}\right), \label{eqn:35}
\end{align}
\end{footnotesize}
\begin{align}
E^{\text{fly}}_k(\{\tau_0^{\text{fly}}[n]\})= &\sum_{n=1}^N \eta{P} h_k(\bold{q}^{\text{fly}}[n])  \tau_{0}^{\text{fly}}[n]. \label{eqn:36}
\end{align}

Based on \eqref{eqn:33}, \eqref{eqn:34}, \eqref{eqn:35}, and \eqref{eqn:36}, the uplink common throughput maximization problem is reformulated as follows, in which the optimization variables are $\{\varsigma_k\}$, $\{Q^{\text{hover}}_k\}$, $\{\tau_\omega\}$, $\{\tau_k^{\text{fly}}[n]\}$, $\{{Q}^{\text{fly}}_k[n]\}$, and $R$.
\begin{align}
\text{(P3)}&:\max ~ R\nonumber\\
\mathrm{s.t.}~& \widetilde{R_k}(\varsigma_k,Q^{\text{hover}}_k) + R^{\text{fly}}_k(\{\tau_k^{\text{fly}}[n],{Q}^{\text{fly}}_k[n]\}) \ge R,\forall k\in\mathcal{K}\nonumber\\
~&\varsigma_k Q_k^{\text{hover}}+\sum_{n=1}^N \tau_k[n]Q_k^{\text{fly}}[n] \le \widetilde{E_k}(\{\tau_\omega\})+ E^{\text{fly}}_k(\{\tau_0^{\text{fly}}[n]\})\nonumber\\
&~~~~~~~~~~~~~~~~~~~~~~~~~~~~~~~~~~~~~~~~~~~~~~~~~~~~~,\forall k\in\mathcal{K}\nonumber\\
~&\varsigma_k \ge 0,\forall k\in\mathcal{K},\tau_\omega \ge 0,\forall \omega\in\{1,\ldots,\Omega^{(\boldsymbol\mu^{\text{opt}})}\}\label{zeta}\\
~&Q^{\text{hover}}_k\ge 0, {Q}^{\text{fly}}_k[n]\ge 0,\forall k\in\mathcal{K}, n\in\mathcal{N}\label{QKQN}\\
~&\sum_{k=1}^{K} \varsigma_k+\sum_{\omega=1}^{\Omega^{(\boldsymbol\mu^{\text{opt}})}} \tau_\omega= T-T^{\text{fly}}\label{zetasum}\\
&\tau_k^{\text{fly}}[n]\ge 0,\forall n\in\mathcal{N}, k \in \mathcal{K}\cup\{0\}\\
&\sum_{k=0}^K \tau^{\text{fly}}_k[n] = \delta,\forall n \in\mathcal{N}.\label{tausum}
\end{align}
Note that although problem (P3) is non-convex due to the coupling of some variables (e.g., $\varsigma_k$ and $Q^{\text{hover}}_k$). Fortunately, via a change of variables (e.g., by introducing $E_k^{\text{hover}} = \varsigma_k Q^{\text{hover}}_k$), we can transform problem (P3) into a convex optimization problem, which can thus be solved via standard convex optimization techniques. By combining the optimal solution to (P3) together with the successive hover-and-fly trajectory in \eqref{eqn:trajectory:1} and \eqref{eqn:trajectory:2}, the solution to (P1) is finally found.

\begin{remark}\label{Remark:4.1}
It is worth noting that the proposed successive hover-and-fly trajectory design is asymptotically optimal for (P1) when $T\rightarrow \infty$, as the total flying time $T_{\text{fly}}$ becomes negligible as compared to the total hovering time $T-T_{\text{fly}}$. In this case, the obtained uplink common throughput approaches the optimal value of (P2), which serves as the upper bound for that of (P1).
\end{remark}
\subsection{Trajectory Redesign and Transmission Resource Allocation When $T< T_{\text{fly}}$}\label{T<T}
Next, we consider the case with $T< T_{\text{fly}}$, in which the UAV flying trajectory $\{\tilde{\bold{q}}(t)\}_{t=0}^{T_{\text{fly}}}$ based on the TSP solution is no longer feasible since the duration $T$ is not sufficient for the UAV to visit all the ${\Omega^{(\boldsymbol\mu^{\text{opt}})}}+K$ hovering locations. To overcome this problem, we first find the solution to (P1) when $T$ is sufficiently small (i.e., $T\rightarrow0$) such that the UAV can only hover at one single location, and then reconstruct a modified UAV trajectory for the case of $T<T_{\text{fly}}$.

First, when $T\rightarrow0$, the UAV should hover at one single fixed location. Let $\bold{q}$ denote the hovering location of the UAV. Then problem (P1) can be re-expressed as
\begin{align}
\max_{\substack{\{\rho_k,Q_k(t)\},\bold{q}}}& ~ \min_{k\in\mathcal{K}}R_k(\{\rho_k,\bold{q},Q_k(t)\})\nonumber\\
\mathrm{s.t.}~~~&\int_0^T \rho_k(t) Q_k(t) \text{d}t \le \int_0^T E_k(\rho_0(t),\bold{q})\text{d}t, \forall k\in\mathcal K\nonumber\\
~&Q_k(t) \ge 0,\forall k \in\mathcal{K}, t\in\mathcal T\nonumber\\
~&\rho_k(t)\in\{0,1\},\forall k \in\{0\}\cup\mathcal{K}, t\in\mathcal T\nonumber\\
~&\sum_{k=0}^K \rho_k(t)= 1,\forall t\in\mathcal T.\label{Qfixproblem}
\end{align}
Note that under given UAV hovering location $\bold{q}$, it is easy to show that problem (\ref{Qfixproblem}) is equivalent to the common throughput maximization problem in the conventional WPCN \cite{Rui2013WPCN}. In this case, the transmission resource allocation can be obtained optimally by performing a TDMA protocol together with a joint time and power allocation, for which the details can be found in \cite{Rui2013WPCN}. Therefore, the optimal $\bold{q}$ to problem (\ref{Qfixproblem}) can be obtained via a 2D exhaustive search over $[\underline{x},\overline{x}]\times[\underline{y},\overline{y}]$, together with the joint time and power allocation under any given $\bold{q}$. We denote the obtained optimal $\bold{q}$ to problem (\ref{Qfixproblem}) as $\bold{q}_{\text{fix}}$.

With $\bold{q}_{\text{fix}}$ at hand, we then reconstruct the trajectory for problem (P1) as follows by down-scaling the previously obtained traveling path $\{\tilde{\bold{q}}(t)\}_{t=0}^{T_{\text{fly}}}$ for the case of $T=T_{\text{fly}}$ linearly towards the center point $(x_{\text{fix}},y_{\text{fix}},H)$, such that the resulting total flying distance equals $V_{\max}T$.
Accordingly, we can zoom in the trajectory as follows:
\begin{align}
&x^{**}(t)=\hat x(t/\nu)+(1-\nu)(x_{\text{fix}}-\hat x(t/\nu)),\nonumber\\
&y^{**}(t)=\hat y(t/\nu)+(1-\nu)(y_{\text{fix}}-\hat y(t/\nu)),\forall t\in[0,T],\label{TleTT}
\end{align}
where $\nu=T/T_{\text{fly}}<1$ denotes the linear scaling factor. Note that when $T\rightarrow0$, we have $\nu\rightarrow0$, and the above redesigned trajectory reduces to hovering at one single fixed location $\bold{q}_{\text{fix}}$; when $T\rightarrow T_{\text{fly}}$, we have $\nu\rightarrow1$, and the above redesigned trajectory becomes identical to the TSP-based trajectory $\{\tilde{\bold{q}}(t)\}_{t=0}^{T_{\text{fly}}}$.

In this case, the successive hover-and-fly trajectory can be modified as $\{\bold{q}^{**}(t)\}$ with $\bold{q}^{**}(t) = (x^{**}(t),y^{**}(t)),\forall t\in[0,T]$. Accordingly, the transmission resource allocation can be obtained similarly as for (P3) by discretizing the $T$ into $N$ time slots and employing TDMA within each slot for WPT and WIT, respectively. The details of the transmission resource allocation are omitted for convenience.

In summary, we present the overall algorithm with the successive hover-and-fly trajectory for solving (P1) as Algorithm 2 in Table \ref{algorithm:2}, with both the cases of $T\ge T_{\text{fly}}$ and $T<T_{\text{fly}}$.
\begin{table}
\begin{center}
\caption{Algorithm 2 for Solving Problem (P1)} \vspace{0.1cm}
\hrule \vspace{0.3cm}
\begin{itemize}
\item[a)] Solve problem (P2) by Algorithm 1 to find the $\Omega^{(\boldsymbol\mu^{\text{opt}})}+K$ optimal hovering locations $\{{\bold{q}}^{\text{opt}}_{k}\}_{k=1}^{\Omega^{(\boldsymbol\mu^{\text{opt}})}+K}$.
\item[b)] Add a dummy hovering location, namely the $(\Omega^{(\boldsymbol\mu^{\text{opt}})}+K+1)$-th hovering location, and set its distances to all the existing $\Omega^{(\boldsymbol\mu^{\text{opt}})}+K$ hovering locations as 0.
\item[c)] Obtain the desirable traveling path $\{\tilde{\bold{q}}(t)\}_{t=0}^{T_{\text{fly}}}$ by solving the standard TSP problem for the $\Omega^{(\boldsymbol\mu^{\text{opt}})}+K+1$ hovering locations and then remove the two edges associated with the dummy location, where $T_{\text{fly}}$ denotes the total flying time.
\item[d)] If $T\ge T_{\text{fly}}$, then find the optimal hovering time allocation, and transmission resource allocation by solving problem (P3); accordingly, obtain the corresponding trajectory as in Section \ref{TgeTT}.
\item[e)] Otherwise, if $T< T_{\text{fly}}$, then obtain the trajectory based on (\ref{TleTT}), and accordingly find the optimal transmission resource allocation.
\end{itemize}
\vspace{0.1cm} \hrule\label{algorithm:2}
\end{center}
\end{table}
\begin{remark}\label{Remark:4.2}
It is worth further discussing the special case with $K=2$ users to provide more design insights. As mentioned in Remark \ref{twousers}, at the optimal solution to (P2) in this case, there are in general three or four hovering locations, which are all located above the line connecting the two users. Therefore, the corresponding successive hover-and-fly trajectory always flies above the line between the two users, and the total flying duration for visiting all these hovering locations is $T_{\text{fly}} = D/V_{\text{max}}$, with $D$ denoting the distance between the two users. Due to the symmetry between the two users, it can be shown via contradiction that in the case of $T \ge T_{\text{fly}}$, the successive hover-and-fly trajectory and the correspondingly obtained transmission resource allocation are indeed the globally optimal solution to problem (P1). Nevertheless, such a result does not hold in the general scenario with $K > 2$ users.
\end{remark}
\section{Alternating Optimization Based Solution to Problem (P1)}

In this section, we propose an alternative solution to problem (P1) based on the technique of alternating optimization, which optimizes the UAV trajectory and the transmission resource allocation in an alternating manner towards a locally optimal solution. Towards this end, we reformulate problem (P1) by discretizing the whole period $\mathcal T$ with duration $T$ into a finite number of $\widetilde{N}$ time slots, each with duration $\widetilde{\delta}=T/\widetilde{N}$. Note that the duration $\widetilde{\delta}$ is chosen to be sufficiently small, such that we can assume the UAV's location is approximately unchanged during each slot $n$, which is denoted as $(x[n],y[n],H), n\in\widetilde{\mathcal N}\triangleq\{1,...,\widetilde{N}\}$. Similarly as for problem (P3), we divide each slot $n$ into $(K+1)$ sub-slots, with the first sub-slot with duration $\tau_0[n] \ge 0$ for downlink WPT, and the $k$-th sub-slot with duration $\tau_k[n] \ge 0$ for user $k$'s uplink WIT with transmit power $Q_k[n],k\in\mathcal K$. Here, we have $\sum_{k=0}^K \tau_k[n]= \widetilde{\delta}, \forall n \in\widetilde{\mathcal{N}}$. Accordingly, the uplink achievable data rate and the harvested energy at each user $k\in\mathcal K$ at slot $n\in\widetilde{\mathcal N}$ are respectively given as
\begin{align}
\widehat{r_k}(\tau_k[n],{Q}_k[n],\bold{q}[n])= &\tau_{k}[n]{\rm{log_2}}\left(1+\frac{h_k(\bold{q}[n]){Q}_k[n]}{\sigma^2}\right), \label{scp1}\\
\overline{E_k}(\tau_0[n],\bold{q}[n])= & \eta{P} h_k(\bold{q}[n])  \tau_{0}[n]. \label{scp2}
\end{align}
Accordingly, problem (P1) can be reformulated as
\begin{align}
&\text{(P4)}:\max_{\substack{\{\tau_k[n],{Q}_k[n],\bold{q}[n]\}}} ~ \min_{k\in\mathcal K}\frac{1}{T}\sum_{n=1}^{\widetilde{N}}\widehat{r_k}(\tau_k[n],{Q}_k[n],\bold{q}[n])\nonumber\\
&\mathrm{s.t.}~\sum_{n=1}^{\widetilde{N}} \tau_k[n]Q_k[n] \le \sum_{n=1}^{\widetilde{N}}\overline{E_k}(\tau_0[n],\bold{q}[n]), \forall k \in\mathcal{K} \label{con53}\\
&~~~~\lVert\bold{q}[n+1]-\bold{q}[n]\rVert^2\le V_{\max}^2\widetilde{\delta}^2, \forall n \in\{1,...,\widetilde{N}-1\}\label{con43}\\
&~~~~\sum^{K}_{k=0}\tau_k[n]\le\widetilde{\delta},\forall n\in\mathcal {\widetilde{N}}\label{scptau}
\end{align}
where the constraints in (\ref{con43}) correspond to the discretized version of the UAV's maximum speed constraint in (\ref{speed}). In the following, we optimize each of the UAV trajectory $\{\bold{q}[n]\}$ and the transmission resource allocation $\{\tau_k[n],{Q}_k[n]\}$ for (P4), respectively, by assuming the other one is given.

First, we optimize the UAV trajectory $\{\bold{q}[n]\}$ under any given $\{\tau_k[n]\}$ and $\{Q_k[n]\}$. In this case, problem (P4) is not a convex optimization problem with respect to $\{\bold{q}[n]\}$, as the rate function in the objective function and the energy function at the right-hand-side (RHS) of (\ref{con53}) are both non-concave. To tackle this issue, we propose an efficient algorithm by using the SCP technique, which updates the UAV trajectory $\{\bold{q}[n]\}$ in an iterative manner by transforming the non-convex problem into a convex approximate problem. Suppose that $\{\bold{q}^{(0)}[n]\}=\{(x^{(0)}[n],y^{(0)}[n])\}$ denotes the initial UAV trajectory and $\{\bold{q}^{(i)}[n]\}=\{(x^{(i)}[n],y^{(i)}[n])\}$ corresponds to the obtained UAV trajectory after iteration $i\ge1$. Then we have the lower bounds for $\overline{E_k}(\tau_0[n],\bold{q}[n])$ and $\widehat{r_k}(\tau_k[n],{Q}_k[n],\bold{q}[n])$ in the following lemma.
\lemma\label{lemma4.1} Under any given UAV trajectory $\{\bold{q}^{(i)}[n]\}$, it follows that
\begin{align}
\overline{E_k}(\tau_0[n],\bold{q}[n])&\ge \overline{E_k}^{(i)}(\tau_0[n],\bold{q}[n]),\label{eqn:hatEk}\\
\widehat{r_k}(\tau_k[n],{Q}_k[n],\bold{q}[n])&\ge \widehat{r_k}^{(i)}(\tau_k[n],{Q}_k[n],\bold{q}[n])\nonumber\\
&~~~~~~~~~~~~~~,\forall k\in\mathcal K, n\in\mathcal N,\label{eqn:47}
\end{align}
where
\begin{align}
\overline{E_k}^{(i)}(\tau_0[n],\bold{q}[n])&\triangleq \frac{2\eta\beta_0P\tau_0[n]}{H^2+\lVert\bold{q}^{(i)}[n]-\bold{w}_k\rVert^2}\nonumber\\
&-\frac{\eta\beta_0P\tau_0[n](H^2+\lVert\bold{q}[n]-\bold{w}_k\rVert^2)}{(H^2+\lVert\bold{q}^{(i)}[n]-\bold{w}_k\rVert^2)^2},\label{equ:Ei}
\end{align}
\begin{align}
&\widehat{r_k}^{(i)}(\tau_k[n],{Q}_k[n],\bold{q}[n])\triangleq \tau_{k}[n]{\rm{log_2}}\left(1+\frac{Q_k[n]\gamma}{H^2+S_k^{(i)}[n]}\right)\nonumber\\
&~~~-\frac{\gamma Q_k[n]\tau_k[n]\log_2e(S_k[n]-S^{(i)}_k[n])}{(H^2+S^{(i)}_k[n])^2+\gamma Q_k[n]\tau_k[n](H^2+S^{(i)}_k[n])},\label{eqn:48}
\end{align}
with $e$ defining the Euler's number, $S_k^{(i)}[n]\triangleq \lVert\bold{q}^{(i)}[n]-\bold{w}_k\rVert^2$, and $S_k[n]\triangleq \lVert\bold{q}[n]-\bold{w}_k\rVert^2$. Notice that $\overline{E_k}^{(i)}(\tau_0[n],\bold{q}[n])$ and $\widehat{r_k}^{(i)}(\tau_k[n],{Q}_k[n],\bold{q}[n])$ are both concave functions with respect to $\bold{q}[n]$.  The inequalities in (\ref{eqn:hatEk}) and (\ref{eqn:47}) are tight for $\bold{q}[n]=\bold{q}^{(i)}[n]$, i.e.,
\begin{align}
\overline{E_k}(\tau_0[n],\bold{q}^{(i)}[n])&= \overline{E_k}^{(i)}(\tau_0[n],\bold{q}^{(i)}[n]),\\
\widehat{r_k}(\tau_k[n],{Q}_k[n],\bold{q}^{(i)}[n])&= \widehat{r_k}^{(i)}(\tau_k[n],{Q}_k[n],\bold{q}^{(i)}[n])\nonumber\\
&~~~~~~~~~~~~,\forall k\in\mathcal K, n\in\mathcal N.
\end{align}
\begin{IEEEproof}
See Appendix \ref{Appendix:4}.
\end{IEEEproof}

Based on Lemma \ref{lemma4.1}, at each iteration $i+1$, we optimize over $\bold{q}[n]$ by replacing $\overline{E_k}(\tau_0[n],\bold{q}[n])$ and $\widehat{r_k}(\tau_k[n],{Q}_k[n],\bold{q}[n])$ in problem (P4) with their respective lower bounds $\overline{E_k}^{(i)}(\tau_0[n],\bold{q}[n])$ in (\ref{equ:Ei}) and $\widehat{r_k}^{(i)}(\tau_k[n],{Q}_k[n],\bold{q}[n])$ in (\ref{eqn:48}), respectively, with the obtained UAV trajectory $\{\bold{q}^{(i)}[n]\}$ at the previous iteration $i$. More specifically, the UAV trajectory is updated as
\begin{align}
\bold{q}^{(i+1)}[n]=&\arg\max_{\{\bold{q}[n]\}}~\min_{k\in\mathcal{K}}\frac{1}{T}\sum^{\widetilde{N}}_{n=1}\widehat{r_k}^{(i)}(\tau_k[n],{Q}_k[n],\bold{q}[n])\nonumber\\
\mathrm{s.t.}~&\sum_{n=1}^N \tau_k[n]Q_k[n] \le \sum_{n=1}^N\overline{E_k}^{(i)}(\tau_0[n],\bold{q}[n]), \forall k\in\mathcal K\nonumber\\
&\text{(\ref{con43})}.\label{con50}
\end{align}
Note that for problem (\ref{con50}), the function $\widehat{r_k}^{(i)}(\tau_k[n],{Q}_k[n],\bold{q}[n])$ is concave with respect to $\bold{q}[n]$, and the constraints are convex. As a result, problem (\ref{con50}) is a convex optimization problem, and thus can be optimally solved by standard convex optimization techniques such as the interior point method\cite{Boyd2004}. Notice that the objective function in problem (\ref{con50}) serves as a lower bound for that in problem (P4) (see Lemma \ref{lemma4.1}). Therefore, it follows that after each iteration $i\ge 1$, the objective value of problem (P4) achieved by $\{\bold{q}^{(i)}[n]\}$ is no smaller than that achieved by $\{\bold{q}^{(i-1)}[n]\}$ in the previous iteration $(i-1)$. As the optimal value of problem (P4) is bounded from above, the SCP-based design in (\ref{con50}) converges to a locally optimal trajectory solution under given transmission resource allocation $\{\tau_k[n]\}$ and $\{Q_k[n]\}$.

Next, we optimize the transmission resource allocation $\{\tau_k[n]\}$ and $\{{Q}_k[n]\}$ under given UAV trajectory $\{\bold{q}[n]\}$. Although this problem is non-convex, similarly as for problem (P3), we can transform it into a convex optimization problem via change of variables, which can thus be solved  via standard convex optimization techniques\cite{Boyd2004}.

Finally, we optimize over the UAV trajectory $\{\bold{q}[n]\}$ via (\ref{con50}) based on the SCP technique and the transmission resource allocation $\{\tau_k[n]\}$ and $\{Q_k[n]\}$ via the convex optimization technique, in an alternating manner. It is worth noting that such an alternating optimization ensures the objective function of problem (P4) to be monotonically non-decreasing after each iteration with all variables updated. As a result, the alternating-optimization-based approach eventually converges to a locally optimal solution to (P4).

It is also worth noting that as problem (P4) is a non-convex optimization problem that generally possesses multiple locally optimal solutions, the performance of the alternating-optimization-based approach for solving (P4) critically depends on the initial point chosen for iteration. In this paper, we choose the successive hover-and-fly trajectory based solution in Section \ref{sec:P1} as the initial point. In this case, the alternating-optimization-based approach in this section can always achieve a common throughput no smaller than the successive hover-and-fly trajectory based design. As the successive hover-and-fly trajectory is a reasonably good heuristic design, this ensures the alternating-optimization based algorithm to achieve a better common throughput after convergence, as will be validated by our numerical results in the next section.

\begin{figure}
  \centering
  \includegraphics[width=7cm]{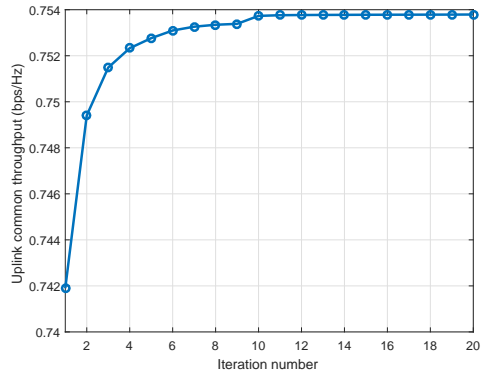}\\
  \caption{Convergence behavior of the alternating-optimization-based approach for solving problem (P4).}\label{fig:77}
\end{figure}

For the purpose of illustration, Fig. \ref{fig:77} shows the convergence behavior of the alternating-optimization-based algorithm for solving problem (P4), where the parameters are set similarly as for Fig. \ref{fig:2} (which will be detailed later) with the time duration $T=4$~s. It is observed that the uplink common throughput (or the objective value of problem (P4)) is monotonically increasing after each iteration, and the iteration converges very fast.

\section{Numerical Results}\label{sec:simulation}

In this section, we present numerical results to validate the performance of our proposed joint trajectory and transmission resource allocation design as compared to the following banchmark scheme.
\begin{itemize}
\item  {\bf Static hovering}: The UAV hovers at a fixed location $\bold{q}$ over the whole period. In this case, the uplink common throughput maximization problem is reduced to problem (\ref{Qfixproblem}), which can be solved efficiently via a 2D exhaustive search over $\bold{q}$ and solving for the transmission resource allocation under any given $\bold{q}$.
\end{itemize}


In the simulation, the UAV flies at a fixed altitude $H=5$~m. The receiver noise power at the UAV is set as $\sigma^2=-80$ dBm. The channel power gain at the reference distance $d_0=1$ m is set as $\beta_0=-30$ dB. The energy harvesting efficiency is set as $\eta=50\%$. The maximum speed of the UAV is $V_{\text{max}}=10$ m/s. The transmit power at the UAV is $P=40$~dBm.
\begin{figure}
  \centering
  \includegraphics[width=7cm]{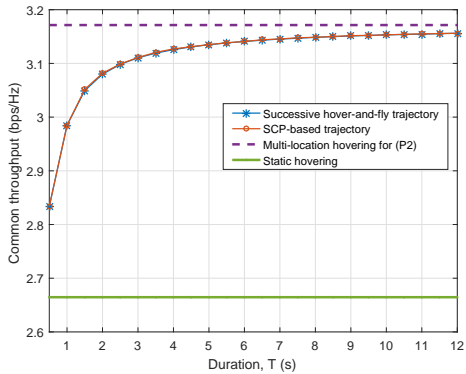}\\
  \caption{Uplink common throughput versus the flight duration $T$ for the case with $K=2$ users and $D=10$~m.}\label{fig:7}
\end{figure}


First, we consider the special case with $K = 2$ users. Fig. \ref{fig:7} shows the uplink common throughput in this case versus the flight duration $T$, in which the distance between the two users is set as $D=10$~m. It is observed that the successive hover-and-fly trajectory has the same performance as the SCP-based trajectory. This is consistent with Remark \ref{Remark:4.2}, which shows that the successive hover-and-fly trajectory is indeed optimal to (P1) when $T \ge T_{\text{fly}} = D/V_{\text{max}}= 1$~s. As a result, in this case, the SCP-based trajectory cannot further improve the performance. Furthermore, it is observed that the proposed successive hover-and-fly trajectory and SCP-based trajectory achieve higher common throughput than the static-hovering benchmark, and the performance gain becomes more substantial when $T$ becomes larger. Last, the two proposed designs are observed to approach the performance upper bound by the multi-location-hovering solution for (P2). This is consistent with Remark \ref{Remark:4.1}.

\begin{figure}
  \centering
  \includegraphics[width=7cm]{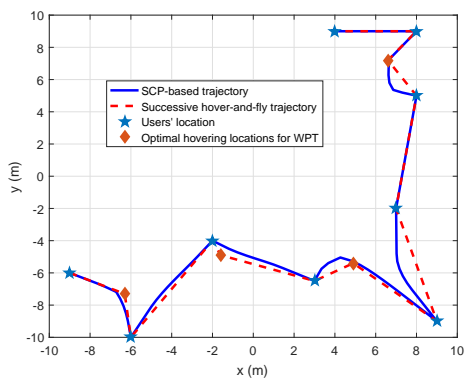}\\
  \caption{System setup for simulation and various trajectories obtained with $T=12$ s.}\label{fig:2}
\end{figure}


Next, we consider a system with $K=9$ ground users that are randomly distributed within a 2D area of $20 \times 20$ m$^2$, as shown in Fig. \ref{fig:2}. For illustration, Fig. \ref{fig:2} also shows the optimal hovering locations for problem (P2), the successive hover-and-fly trajectory as well as the SCP-based trajectory under the 9-user setup, where $T = 12$~s. It is observed that there are $\Omega^{(\boldsymbol\mu^{\text{opt}})}=4$ optimal hovering locations for WPT and a total of $\Omega^{(\boldsymbol\mu^{\text{opt}})} + K =13$ optimal hovering locations for problem (P2).

\begin{figure}
\begin{minipage}[t]{0.5\textwidth}
\centering
    \includegraphics[width=7cm]{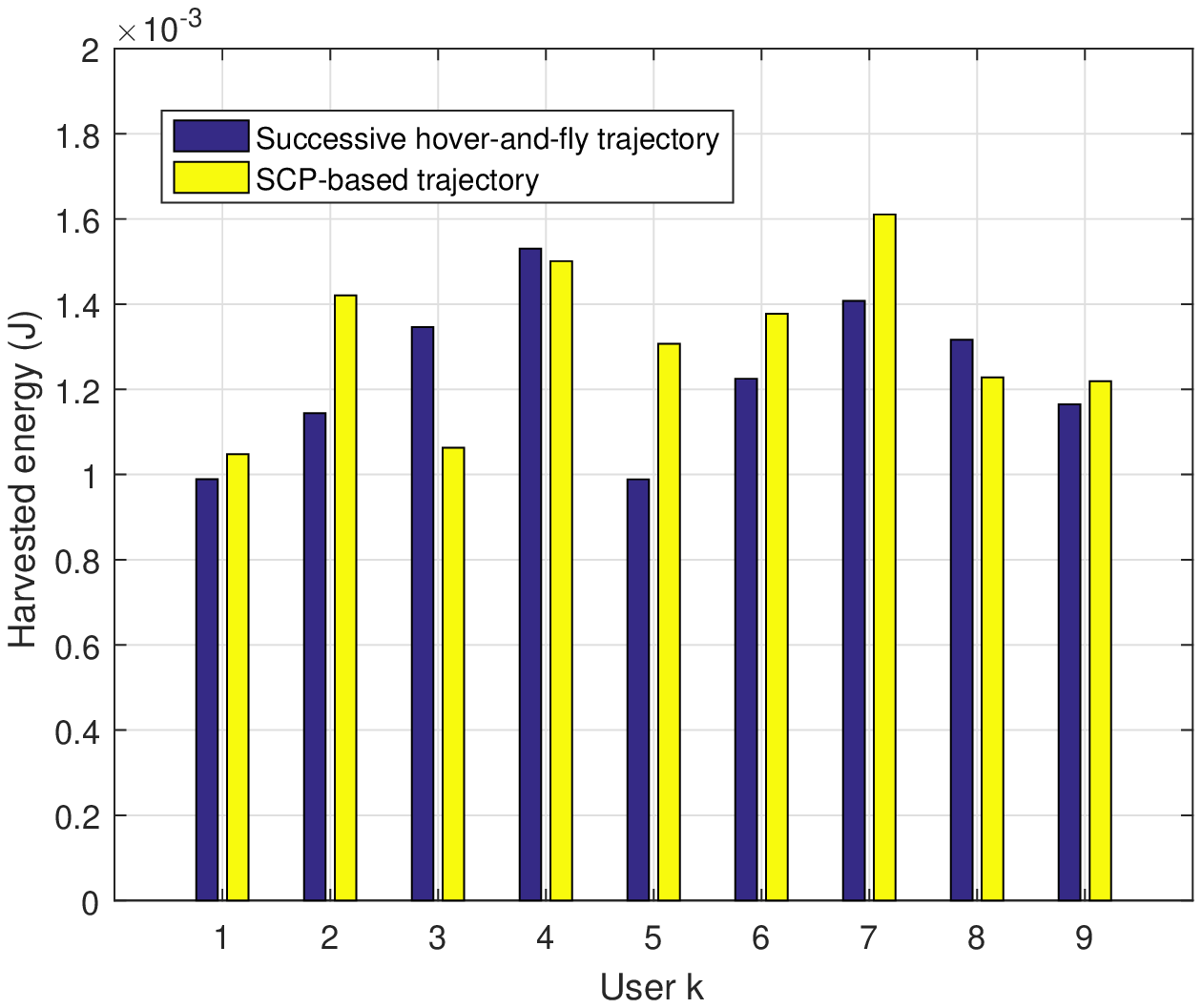}
\caption{The harvested energy at the $K$ users.} \label{fig:5}
\end{minipage}
\begin{minipage}[t]{0.5\textwidth}
\centering
    \includegraphics[width=7cm]{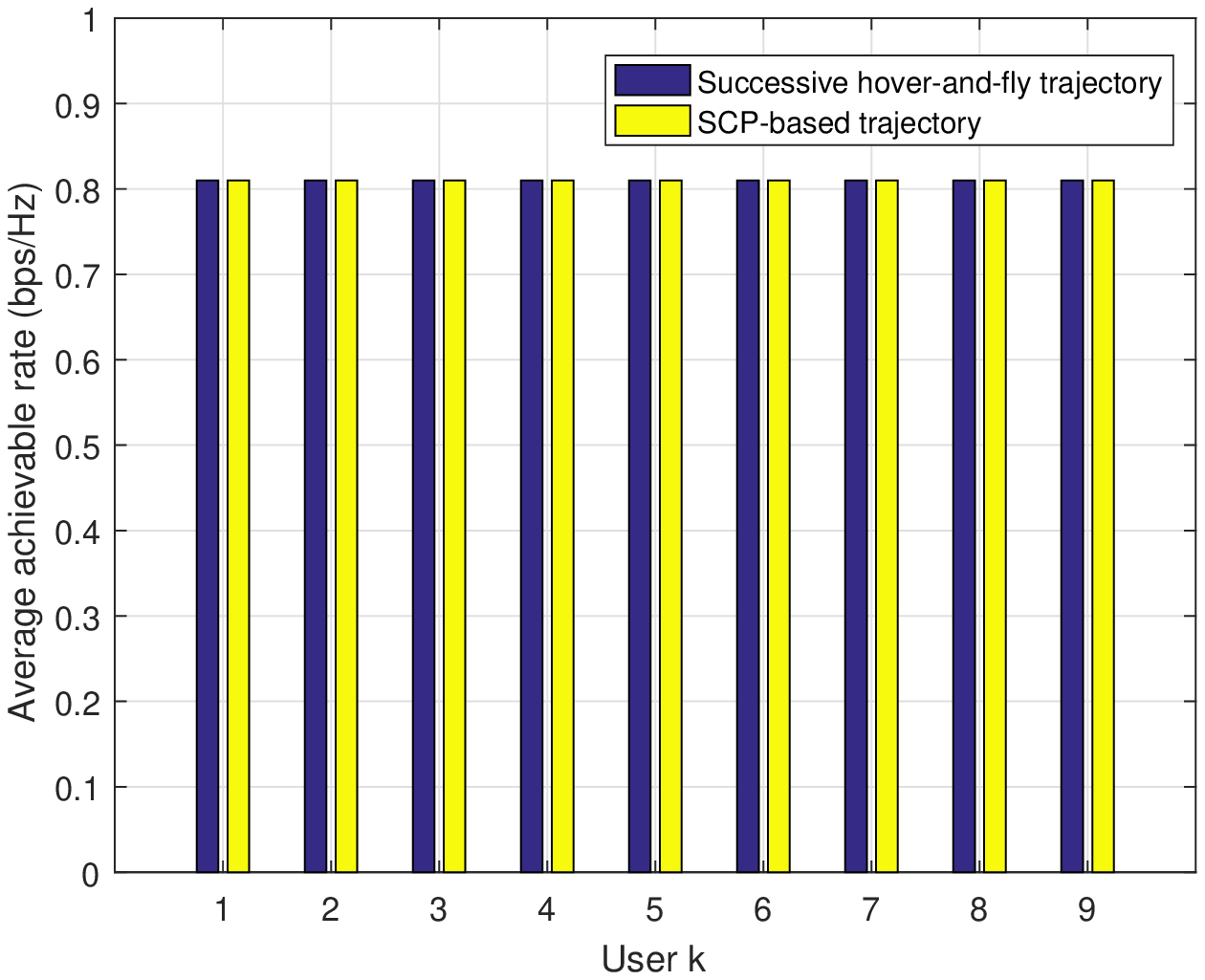}
\caption{The average achievable rate at the $K$ users.} \label{fig:6}
\end{minipage}
\end{figure}

Fig. \ref{fig:5} and Fig. \ref{fig:6} show the harvested energy and the achievable rate of each ground user, respectively, with $T=12$ s. It is observed in Fig. \ref{fig:5} that in both proposed designs, the harvested energy values at different users are generally different. By contrast, it is observed in Fig. \ref{fig:6} that the achievable rates at these users are the same in order to maximize the common throughput. It can be inferred from Figs. \ref{fig:5} and \ref{fig:6} that the trajectory design and the corresponding transmission resource allocation for (P1) can efficiently balance the communication rates among the $K$ users, but generally may lead to unbalanced energy harvesting at these distributed users. This shows the difference between our design versus that in the UAV-enabled multiuser WPT system \cite{JieXuWPT}, in which the harvested energy of all users are designed to be identical.

\begin{figure}
  \centering
  \includegraphics[width=7cm]{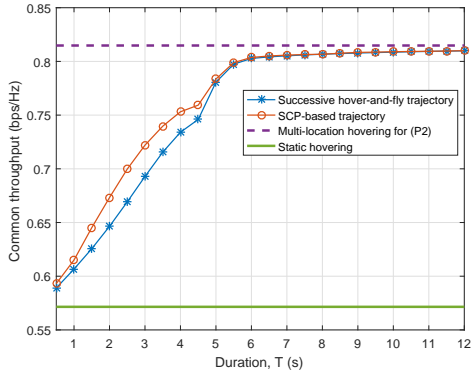}\\
  \caption{Uplink common throughput versus the flight duration $T$.}\label{fig:3}
\end{figure}
Fig. \ref{fig:3} shows the uplink common throughput of the $K$ users versus the flight duration $T$. It is observed that the proposed successive hover-and-fly trajectory and SCP-based trajectory (jointly with the corresponding optimized transmission resource allocation) achieve higher common throughput than the static-hovering benchmark, and the performance gain becomes more substantial when $T$ becomes larger. In particular, the SCP-based trajectory outperforms the successive hover-and-fly trajectory, especially when the flight duration is small. Furthermore, with $T$ being sufficiently large, the successive hover-and-fly trajectory and the SCP-based trajectory are both observed to approach the performance upper bound by the multi-location-hovering solution to (P2) with the UAV's maximum speed constraint ignored. This is consistent with Remark \ref{Remark:4.1}.

%
\section{Conclusion}\label{sec:conclusions}
In this paper, we investigate the common throughput maximization problem in a new UAV-enabled WPCN setup, by jointly optimizing the UAV trajectory and the transmission resource allocation in both downlink WPT and uplink WIT, subject to the UAV's maximum speed constraint and the users' energy neutrality constraints. To solve this challenging problem, we first consider the ideal case without the UAV's maximum speed constraint and solve the relaxed problem optimally. The optimal solution shows that the UAV should successively hover above two sets of optimal ground locations for downlink WPT and uplink WIT, respectively. Next, based on the optimal solution to the relaxed problem, we propose the successive hover-and-fly trajectory and the SCP-based trajectory to solve the problem with the UAV's maximum speed constraint considered. Numerical results showed that the proposed UAV-enabled WPCN achieves near-optimal performance when the flight period is sufficiently large, and significantly enhances the uplink common throughput performance over the conventional WPCN with a static AP located even at the optimal location, by effectively resolving the ``doubly near-far'' fairness issue.

Due to the space limitation, there are several unaddressed  problems in this paper, which are worth investigation in future work. For example, this paper only considered a single UAV serving multiple users, while how to extend to the case with multiple UAVs serving a large number of users over a large area is an interesting open problem. In this case, collaborative downlink energy beamforming and efficient uplink interference control via multi-UAV trajectory design can be employed to enhance the network performance. Besides, this paper maximized the uplink common throughput of multiple users by assuming a fixed UAV mission period. In practice, the UAV mission period should be set by considering various factors such as the communication delay requirements and the battery size of different users, as well as the battery lifetime and power consumption of the UAV. How to design the UAV mission period under the above practical considerations is also an interesting problem worth further study.

\appendix

\subsection{Proof of Proposition \ref{proposition:3.2}}\label{Appendix:2}
Suppose that these $K$ terms are not identical at the optimal $\boldsymbol{\lambda}^{\text{opt}}$ and $\boldsymbol{\mu}^{\text{opt}}$, for which there are in general $K+1$ cases discussed in the following.

In the first case, suppose that $\phi(\{\bar{\bold{q}}_\omega^{(\boldsymbol\mu^{\text{opt}})}\},\{\mu_k^{\text{opt}}\})$ is smaller than any one of $\varphi_k(\bold{w}_k,Q^{(\lambda_k^{\text{opt}},\mu_k^{\text{opt}})},\lambda_k^{\text{opt}},\mu_k^{\text{opt}})$. It follows from Proposition 3.1 that $\rho_0(t) = 0$, $\forall t\in\mathcal T$, i.e., the WPCN does not work in the downlink WPT mode throughout the whole period with duration $T$. In this case, the $K$ users cannot harvest any energy from the UAV, thus leading to a zero common throughput. This solution is thus not optimal.

In the $k$-th one of the other $K$ cases, suppose that $\varphi_k(\bold{w}_k,Q^{(\lambda_k^{\text{opt}},\mu_k^{\text{opt}})},\lambda_k^{\text{opt}},\mu_k^{\text{opt}})$ is smaller than any one of the other $K$ terms (i.e., $\phi(\{\bar{\bold{q}}_\omega^{(\boldsymbol\mu^{\text{opt}})}\},\{\mu_k^{\text{opt}}\})$ and $\varphi_j(\bold{w}_j,Q^{(\lambda_j^{\text{opt}},\mu_j^{\text{opt}})},\lambda_j^{\text{opt}},\mu_j^{\text{opt}})$'s, $\forall j\in\mathcal K$, $j\neq k$). It follows from Proposition 3.1 that $\rho_k(t) = 0$, $\forall t\in\mathcal T$, i.e., the WPCN does not work in the uplink WIT mode for user $k$ throughout the whole period with duration $T$. In this case, the throughput for user $k$ is zero, thus resulting in a zero common throughput. This solution thus cannot be optimal as well.

By combining all the $K+1$ cases in the above, in order for the common throughput to be non-zero, it must hold that $\phi(\bar{\bold{q}}_\omega^{(\boldsymbol\mu^{\text{opt}})},\{\mu_k^{\text{opt}}\}) = \varphi_k(\bold{w}_k,Q^{(\lambda_k^{\text{opt}},\mu_k^{\text{opt}})},\lambda_k^{\text{opt}},\mu_k^{\text{opt}}),\forall k\in\mathcal K,\omega\in\{1,\dots,\Omega^{(\boldsymbol\mu^{\text{opt}})}\}$. Therefore, Proposition \ref{proposition:3.2} is proved.

\subsection{Proof of Lemma \ref{lemma4.1}}\label{Appendix:4}
Define function $g_1(z)=\frac{\eta P\beta_{\rm 0} \tau_0[n]}{H^2+z}$ and $g_2(z)=\tau_{k}[n]{\rm{log_2}}\left(1+\frac{Q_k[n]\gamma}{H^2+z}\right)$, which are both convex with  respect to $z\ge0$. As the first-order Taylor expansion of a convex function is a global under-estimator of the function values, for any given $z_0\ge0$, it follows that $g(z)\ge g(z_0)+g'(z_0)(z-z_0)$, or equivalently,
\begin{align}
&\frac{\eta P\beta_{\rm 0} \tau_0[n]}{H^2+z}\ge\frac{\eta P\beta_{\rm 0} \tau_0[n]}{H^2+z_0}-\frac{\eta P\beta_{\rm 0} \tau_0[n]}{(H^2+z_0)^2}(z-z_0),\label{scpproof1}\\
&\tau_{k}[n]{\rm{log_2}}\left(1+\frac{Q_k[n]\gamma}{H^2+z}\right)\ge\tau_{k}[n]{\rm{log_2}}\left(1+\frac{Q_k[n]\gamma}{H^2+z_0}\right)\nonumber\\
&~~~~~-\frac{\gamma Q_k[n]\tau_k[n]\log_2e}{(H^2+z_0)^2+\gamma Q_k[n]\tau_k[n](H^2+z_0)}(z-z_0).\label{scpproof2}
\end{align}
For any given $k\in\mathcal K, n\in\mathcal N,$ and $i\ge0$, by substituting $z=\lVert\bold{q}[n]-\bold{w}_k\rVert^2$ and $z_0=\lVert\bold{q}^{(i)}[n]-\bold{w}_k\rVert^2$ into (\ref{scpproof1}) and (\ref{scpproof2}), then (\ref{equ:Ei}) and (\ref{eqn:48}) follow, respectively. Furthermore, note that the equality holds for (\ref{scpproof1}) and (\ref{scpproof2}) for $z=z_0$, and therefore, the equality in (\ref{eqn:47}) holds. Therefore, Lemma \ref{lemma4.1} is proved.

\begin{small}

\end{small}

\end{document}